# Dynamic Heterogeneity, Cooperative Motion, and Johari-Goldstein $\beta$-Relaxation in a Metallic Glass-Forming Material Exhibiting a Fragile to Strong Transition

Hao Zhang[1†], Xinyi Wang[1], Hai-Bin Yu[2], Jack F. Douglas[3†]

[1] Department of Chemical and Materials Engineering, University of Alberta, Edmonton, Alberta, Canada, T6G 1H9

[2] Wuhan National High Magnetic Field Center, Huazhong University of Science and Technology, Wuhan, Hubei, China, 430074

[3] Material Measurement Laboratory, Material Science and Engineering Division, National Institute of Standards and Technology, Gaithersburg, Maryland, USA, 20899

*Corresponding authors: hao.zhang@ualberta.ca; jack.douglas@nist.gov

**Abstract**

We investigate the Johari-Goldstein (JG) $\beta$-relaxation process in a model metallic glass-forming (GF) material ($Al_{90}Sm_{10}$), previously studied extensively by both frequency-dependent mechanical measurements and simulation studies devoted to equilibrium properties, by molecular dynamics simulations based on validated and optimized interatomic potentials with the primary aim of better understanding the nature of this universal relaxation process from a dynamic heterogeneity (DH) perspective. The present relatively low temperature and long-time simulations reveal a direct correspondence between the JG $\beta$-relaxation time $\tau_{JG}$ and the lifetime of the mobile particle clusters $\tau_M$, defined as in previous DH studies, a relationship dual to the corresponding previously observed relationship between the $\alpha$-relaxation time $\tau_\alpha$ and the lifetime of immobile particle clusters $\tau_{IM}$. Moreover, we find that the average diffusion coefficient $D$ nearly coincides with $D_{Al}$ of the smaller atomic species (Al), and that the 'hopping time' associated with $D$ coincides with $\tau_{JG}$ to within numerical uncertainty, both trends being in accord with experimental studies. This indicates that the JG $\beta$-relaxation is dominated by the smaller atomic species and the observation of a direct relation between this relaxation process and rate of molecular diffusion in GF materials at low temperatures where the JG $\beta$-relaxation becomes the prevalent mode of structural relaxation. As an unanticipated aspect of our study, we find that $Al_{90}Sm_{10}$ exhibits fragile-to-strong (FS) glass-formation, as found in many other metallic GF liquids, but this fact does not greatly alter the geometrical nature of DH in this material and the relation of DH to dynamical properties. On the other hand, the temperature dependence of the DH and dynamical properties, such as the structural relaxation time, can be significantly altered from 'ordinary' GF liquids.

**Introduction**

Relaxation in glass-forming (GF) liquids occurs as a hierarchical process, both in terms of the timescales and length scales of the separate relaxation processes involved, and these distinct relaxation processes have different significance for the observable properties of this broad class of materials. Moreover, the character of these relaxation processes changes greatly with temperature, $T$. Specifically, there is only a single structural relaxation process at elevated $T$ where the properties of fluids exhibit Arrhenius relaxation, and the mass diffusion coefficient $D$ of the molecules and the shear viscosity $\eta$ of the fluid conform to the properties of simple uniform liquids in which the Stokes-Einstein relation between $D$ and $\eta$ is obeyed, even at the scale of molecules, which is frankly remarkable. At a temperature $T$ below a definite onset temperature $T_A$ for non-Arrhenius relaxation, however, a new and general mode of relaxation arises, the so-called $\alpha$-relaxation process - this term reflecting its primary significance for understanding the relaxation properties of everyday fluids. By default, the initial relaxation process at shorter times, occurring typically on a ps timescale in molecular fluids, or a somewhat shorter timescale in metallic GF liquids (see discussion below) and very small molecules such as water, is termed the '$\beta$-relaxation process' [1] (See Fig. 2 below for illustration of the $\alpha$-relaxation and $\beta$-relaxation processes in the Al-Sm material). While much of the modeling of GF liquids has emphasized the $\alpha$-relaxation process, which is understandable given the diverse practical applications in which the dramatic change in the magnitude of the associated relaxation time $\tau_\alpha$ is important ($\tau_\alpha$ and shear viscosity $\eta$ can range by more than 15 orders of magnitude in the range of $T$ below $T_A$, but still above the glass transition temperature $T_g$), at which cooled liquids show a pronounced tendency to exhibit non-equilibrium evolution in their properties over long timescales. Betancourt et al. [1] and Giuntoli et al. [2]

discuss the $\alpha$-relaxation and $\beta$-relaxation processes over a large $T$ range in the context of polymer melts, but these relaxation features are observed in molecular and atomic liquids generally.

In the $T$ range below $T_g$, mobility and relaxation do not cease, as one might expect from naïvely extrapolating the dynamical properties of materials from the $T$ regime about $T_g$. Instead, new relaxation processes emerge in this 'glass' regime where $\tau_\alpha$ can often be taken as effectively infinite from a practical perspective, [3] and these distinct relaxation processes, which are not even directly apparent in the intermediate scattering function, [1] *dominate* diffusion and relaxation and other material properties in the 'glass' state where these materials can be described rheologically as being 'solids', regardless of how we prefer to think about these materials theoretically. In particular, the Johari-Goldstein (JG) $\beta$-relaxation process in the glass regime then 'replaces' the $\alpha$-relaxation process as the fundamental process in relation to understanding the properties of glass materials, and understandably the engineering community has devoted a huge amount of effort aimed at understanding JG $\beta$-relaxation process in metallic, polymeric and other types of GF materials in relation to understanding basic engineering significant material properties such as toughness, brittleness, impact resistance, hardness, etc. We review various ideas about the JG $\beta$-relaxation process in our discussion below, and in our companion paper on the 'fast' dynamics of the present material [4], but at this point we simply note that while this relaxation process has been 'studied to death' experimentally, there is still little scientific agreement about its nature in terms of the type of molecular processes involved. This situation is clearly an impediment in the design of diverse materials for engineering and medical use.

From our brief overview above, the apparent nonequilibrium effects inherently complicate the study of the JG $\beta$-relaxation process and, correspondingly, much of what we know about the JG $\beta$-relaxation process derives from theoretical speculation (Of course, some of the ideas put forward before to explain the JG $\beta$-relaxation process are quite interesting and inspiring to our computational investigation.), informed by various experimental correlations. This at least provides a phenomenological framework to guide material design based on an Edisonian approach to material development. It is now clear that the JG $\beta$-relaxation process is a universal feature of GF materials and that this process exhibits strong correlations with the $\alpha$-relaxation process [5,6] so we may have some hope to understand the JG $\beta$-relaxation process based on our growing understanding of the $\alpha$-relaxation process and the 'fast' $\beta$-relaxation processes (We refer here to the $\beta$-relaxation process discussed above, where the term 'fast' refers to the fact that this relatively high frequency relaxation process occurs on a much shorter timescale than the JG $\beta$-relaxation process $\tau_{JG}$ that is prevalently observed in materials in their glass state. [1]), based on numerous previous simulation studies in GF liquids incipient to forming a glass. We now know from extensive simulation studies that the slowing down of the dynamics of GF liquids, and the corresponding increase of the activation energy for relaxation and diffusion generally involves the parallel growth of 'dynamic heterogeneities' (DH) taking the form of dynamic polymeric structures involving particles having relatively low and high mobility, respectively. This change in mobility is relative to the expectations of Brownian particles and the geometrical form of these heterogeneities is remarkably insensitive to the material composition, similar DH being observed both in polymeric [7] and metallic GF liquids [8], as well as all other GF liquids studied to date. Recent simulations by Yu et al. [9,10] have indicated that the JG $\beta$-relaxation process may be associated with a specific type of DH involving relatively mobile particles exhibiting collective motion in the exchange motion in the form of atomic strings, a form of DH repeatedly observed fluids in the *T* range well above $T_g$, but still below $T_A$, where the fluid is still in equilibrium and at the same time non-Arrhenius structural relaxation, shear viscosity and mass diffusion also prevail.

Given our extensive previous work on characterizing DH in GF metallic and polymer liquids [7,8,11], we decided to examine the nature of DH in relation to the JG *β*-relaxation process to determine if this phenomenon fits in with the previously developed framework for defining different kinds of DH and their characteristic lifetimes in the fluid regime. In short, we find below that $\tau_\alpha$ corresponds to the lifetime of dynamic *immobile particle clusters* $\tau_{IM}$ and $\tau_{JG}$ corresponds to the lifetime of the *mobile particle clusters,* as defined in our previous work. [7,11] This leads to the simple, intuitive, and arguably almost 'obvious' picture that the mobile particle domains dominate diffusion, while the immobile particles domains dominate the rate of structural relaxation in GF materials. We have previously shown that the mobile particle clusters, as defined by a now standard algorithm, are themselves clusters of more fundamental clusters exhibiting string-like collective motion [8] so our new results are in qualitative accord with previous observations of Sun et al. [10] suggesting a link between JG *β*-relaxation and string-like collective motion, although our specific method of defining the 'strings' is somewhat different. Technical matters relating to the definition of these clusters and other conventionally defined measures DH, along with the interrelation between these DH types and significant dynamical properties of GF liquids, are discussed below.

Our investigation of DH relation to the JG *β*-relaxation process must confront a fundamental problem that we admit at the outset. It is difficult, and sometimes not possible, to fully equilibrate our simulated material in the low temperature range where the JG *β*-relaxation process is clearly discernible in mechanical measurements, and nonequilibrium effects are also prevalent in experimental systems in the glass regime. Accordingly, we must adopt a pragmatic metric for judging the 'applicability' of our modeling, as in all previous studies of the properties of GF materials based on molecular dynamics (MD). We 'validate' our simulations by comparing to the properties of the corresponding real material under the same nominal thermodynamic conditions. If we want to study GF materials by simulation, we must step off the stable ground of carefully equilibrated property calculations that have been our primary focus in the past.

Previous work on Al-Sm metallic GF materials have adopted the same practical approach just described. Nonetheless, such MD simulations, based on the same density functional-derived interatomic potential that we utilize, along with a similar cooling history can reproduce very well the JG $\beta$-relaxation process that is observed rheological measurements on Al-Sm glasses. Notably, this important validation was performed previously by Sun et al. [10] Many of the other properties of this model glass forming liquid have been similarly empirically validated based on the same interatomic potential and for the same or similar system size, cooling protocol, etc. so we make no claim with regard to novelty with respect to this aspect of our study. However, this extensive previous validation work makes this particular metallic glass attractive for a study aimed at understanding the fundamental nature of the JG $\beta$-relaxation process, and other aspects of GF liquids that are rather imperfectly understood at the present time.

A important feature of the JG $\beta$-relaxation process in the Al-Sm material is that this relaxation process is exceptionally well-separated in timescale from the $\alpha$-relaxation process, which is normally observed in the frequency domain as just a shoulder on the α-relaxation peak in metallic glass materials, and the material also appears exceptionally resistant to crystallization. These fortuitous properties have been essential in the effective study of the JG $\beta$-relaxation process by MD simulation. This particular material then provides us with an exceptional opportunity to investigate the JG $\beta$-relaxation process from a dynamic heterogeneity perspective. In the present work, we especially focus on the JG $\beta$-relaxation process, which is just one of a whole 'zoology' of 'fast dynamics' processes relevant to understanding relaxation and diffusion in GF materials. A companion paper [4] explores some of these other apparently universal dynamical features, such as the Boson peak, the amplitude of the 'fast' $\beta$-relaxation process, etc.

Although our initial goal was to elucidate the nature of the JG $\beta$-relaxation process, we soon realized after performing our simulations and associated analyses on this model GF liquid that it exhibits a *completely differen*t type of glass formation than anything we had ever observed before in our simulations of either metallic or polymeric GF liquids. In particular, we observe so-called 'fragile-to-strong' (FS) glass formation, which we later found to be actually rather common in metallic glass materials [12] and, indeed, many other GF liquids, especially those in which the molecules exhibit a tendency to form dynamic network structures that naturally arise from the presence of highly-directional interactions between the atoms or molecules. We discuss this phenomenon extensively in Appendices A and B since our primary focus is on the nature of the JG $\beta$-relaxation process. This choice of paper organization is further motivated by the fact that the DH that we observe in this GF liquid is essentially the same as found in our previous studies of 'ordinary' GF liquids. This is further evidence of the universality of the DH phenomenon in GF liquids broadly.

Although the occurrence of a fragile-to-strong transition was initially an unwanted complication in our investigation of the JG $\beta$-relaxation process, we soon realized that this situation afforded opportunities to ask some new and broad questions about GF liquids. For example, does the previously established relation between the change of the activation free energy for diffusion and change of the average length of particle clusters exhibiting string-like collective motion exist in these GF liquids? A conspicuous feature of fluids exhibiting a FS transition is that the standard Vogel-Fulcher-Tammann (VFT) relation, [13-15] empirically describing the $T$ dependence of $\tau_\alpha$, $\eta$ and $D$ simply does not apply over a large $T$ range to fluids exhibiting this type of glass-formation (We again refer the reader to the Appendices and our development below with regard to the specific deviations from the VFT relation). The strikingly different phenomenology of GF fluids showing FS glass-formation evidently provides an opportunity to test existing theoretical models of glass-formation. We find that the string model of glass-formation [7,16] continues to describe the $T$ dependence of $D$ very well in our Al-Sm alloy over the entire range of glass-formation that we can investigate, as in previous studies of polymeric [7,16] and metallic GF liquids [8] exhibiting 'ordinary' glass-formation. We are not aware of any other physical model that can describe both of these distinct classes of GF materials in a unified way, nor even any effective correlative scheme devoid of any theoretical rationalization. We believe that the theoretical study of the FS transition promises to greatly expand our understanding of GF liquids and in Appendices A and B we provide a discussion of many aspects of this type of glass-formation since this is a problem of significant independent interest.

Given the situation just described, we decided to thoroughly investigate dynamic heterogeneity, collective motion, and the string model of glass-formation, exactly following the methods utilized before in characterizing dynamic heterogeneity in a Cu-Zr metallic glasses[8] of different compositions, and for coarse-grained GF polymer liquids. [7] Since we have characterized our methodology for characterization of dynamic heterogeneity in GF liquids in numerous previous studies, we concentrate on just reporting results on the various standard measures of dynamic heterogeneity. Our brevity in discussing these results is justified by the fact the geometrical form of the heterogeneous dynamics observed in this material is remarkably similar to previous observations on both metallic glass and polymer GF liquids, although there are changes in the *T* dependence of this dynamic heterogeneity that seems to be characteristic of liquids exhibiting FS glass-formation. [7,8] Changes in the nature of glass-formation in systems that exhibit FS glass-formation in comparison to those exhibiting 'ordinary' glass-formation are discussed extensively in Appendices A and B. Despite some notable differences in the *T* dependence of the DH, many aspects of dynamic heterogeneity, especially geometrical aspects, are remarkably universal. All these background results and methodologies then put us in a good position to study the JG *β*-relaxation process from a dynamic heterogeneity perspective.

## II. Model and Simulation Methods

Classical molecular dynamics simulations were performed to investigate thermodynamic and dynamic properties during fragile to strong transition of $Al_{90}Sm_{10}$ metallic glass-forming alloy using a many-body empirical potential developed by Mendelev et al. [17] This semi-empirical potential takes Finnis-Sinclair form [18], where cohesive energy, elastic modulus, vacancy formation energy, melting point of pure aluminum and a series of Al-rich crystalline compounds with Sm concentration near 10 % were used in the potential parameter fitting procedure. This potential employed in the current study has the advantages of providing excellent reproduction of pure Al properties and the formation energy of a series of Al-rich crystal phases, excellent reproduction of liquid structure of $Al_{90}Sm_{10}$ at *T* = 1273 K and good agreement with ab initio molecular dynamics results so we expect this model to be suitable for simulating glass-formation in this alloy. [17,19]

The liquid sample, containing 28785 Al atoms and 3215 Sm atoms, was initially held at 2000 K for 2.5 ns to reach equilibrium. The liquid then was continuously cooled down with a constant cooling rate of 0.1 K/ns to 200K, and the total simulation time for cooling was 18 μs, an exceptionally long time in comparison to our former simulations of GF liquids. Although an extremely low cooling rate was employed in the current study, the system will not be able to reach *complete* equilibrium at low temperatures where the reciprocal cooling rate becomes longer than the relaxation time. Note the relaxation time determined based on the self-intermediate scattering function below at FS transition temperature, $T_{FS}$, and lambda transition, $T_\lambda$, are 630 ps and 43 ps, respectively, which is orders magnitude shorter than the current cooling rate, suggesting complete equilibrium is expected at these characteristic temperatures under the current simulation conditions.

Periodic boundary conditions were applied in all directions and the isobaric-isothermal ensemble (NPT) was employed where the zero pressure and simulation box size were controlled by the Parrinello-Rahman method[20] and $T$ was maintained by the Nose-Hoover method. [21,22] The MD simulations utilize LAMMPS [23], which was developed at the Sandia National Laboratories. Isothermal heating for an extended period of time was also applied in current study to ensure the system to reach near equilibrium and to allow us to probe kinetic processes that cannot be observed under continuous heating conditions. In the current study, isothermal heating simulations were performed at $T$ = 900 K, 850 K, 800 K, 750 K, 7000 K, 650 K, 600 K, 550 K, 500 K and 450 K. At each $T$, the simulation was conducted for at least 10 ns and up to 0.7 μs. In the Supplemental Information (SI), we show that temporal evolution of potential energy and number density of the system during isothermal heating simulations at low temperatures do not change with time, suggesting a 'near equilibrium' or long-lived metastable condition has been achieved in the time window that we are observing.

Although our simulations are often on the order of a relatively 'long' timescale of μs, this time is still rather relatively short compared with the structural relaxation time $\tau_\alpha$ near $T_g$, which is on the order of a minute in typical GF liquids. This is an inherent limitation of MD simulations, which has largely precluded investigation of the JG $\beta$-relaxation process in previous studies, with a few exceptions. [24,25] As noted above, non-equilibrium effects are also an inherent issue in measurements performed in the glass state where JG $\beta$-relaxation measurements are normally performed. One of the attractive features of the present system, is that the fragile-strong transition and its thermodynamic anomalies can be accessed by MD simulation long before we become near the glass transition where non-equilibrium effects become overtly prevalent in the simulations.

## III. Results and Discussion

### A. Review of Thermodynamic and Structural Properties of Al-Sm Melts

One of the attractive characteristics of the model GF liquid that we study is that its equilibrium thermodynamic, structural and rheological characteristics have been rather thoroughly investigated in the course of validating the interatomic potential, based on these experimental criteria. For example, previous work has shown that the 'structure' of the liquid state, based on the pair correlation function, is well reproduced at $T = 1273$ K. We also expect the potential to provide a realistic description of the material in its glass state as this potential reproduces the short-range order (SRO), represented by Sm-centered motif, predicted by ab initio molecular dynamics simulations (AIMD), [26] Also, a study of devitrification from the glass state based on this potential reproduces metastable structures directly observed by tunneling electron microscopy (TEM) observations, [27] and imaging studies of crystal growth.[28] Another aspect of this model that has been rather exhaustively investigated for this metallic glass is the tendency of the atomic species to form locally icosahedral packed structures in the liquid and the tendency for these domains to form extended polymeric structures upon approaching the glass transition. [19] The usual changes in the liquid structure, as quantified by the static structure factor, and the tendency of local structure formation at the mesoscale in the form of strings of icosahedral clusters, a phenomenon highly prevalent in Cu-Zr and other metallic glasses showing 'ordinary' glass-formation [29,30], occur in our $Al_{90}Sm_{10}$ metallic glass-forming alloy, but the tendency to form long polymeric structures is less prevalent than in the Cu-Zr metallic GF liquid.[26] We conclude from this type of local structural analysis that there does not appear to be anything obviously 'special' about the $Al_{90}Sm_{10}$ metallic GF liquid that can explain the non-standard pattern of the dynamics of this GF material. See Appendix A for further discussion relating to this point.

Recently, there has been intense research aimed at quantifying the nature of this structural organization in metallic GF materials based on the combination of computational and ultrahigh resolution measurement studies that push the limits of each approach because of the long computational times and high spatial resolution required for each method, respectively. [31-35] The general pattern of behavior revealed by these studies is that the organization in metallic glass-forming liquids generally involves an SRO at an atomic scale in which larger 'solute' atoms, like the Sm atoms in our Al-Sm alloy, are 'solvated' by the smaller atomic species to form well-defined clusters in the metallic GF liquid having an approximate icosahedral symmetry, while the structure characterizing the medium range ordering (MRO) corresponds to polymeric structures comprised of the primary icosahedral clusters. The physical situation then greatly resembles the hierarchical assembly of worm-like micelles and amyloid fibers in which compact clusters first form, and then these structures in turn associate to form polymeric assembles that coexist with the smaller clusters and the monomeric structures. [36-41] In line with this type of hierarchical self-assembly process, both measurements and simulations of metallic glass materials have revealed that an additional structural organization at a larger scale than the individual icosahedral clusters occurs in the metallic glass state in which the strings of icosahedra form and these structures in turn form aggregated domains at even a larger scale on the order of a couple of nm. These relatively 'ordered' regions are surrounded by relatively 'loosely-packed' regions [32,34,35] that are largely 'disordered' and lower in density, in addition to cavities devoid of particles altogether. Nanoprobe measurements of the local elasticity of metallic glass materials have provided additional insights into this nanoscale heterogeneity. [42,43] These observations are broadly consistent with the occurrence of liquid-liquid phase separation of the GF liquid at low temperatures into coexisting phases of distinct entropy, a phenomenon investigated in many works on metallic and metallic oxide GF materials, [44-46] organic GF liquids [47-49] and discussed intensively recently in many simulation and experimental studies in connection with understanding the thermodynamic and dynamic properties of water at low $T$. [50-53] We note that there is evidence of structure formation in the form of equilibrium polymeric structures even in the case of some pure metallic species which also corresponding exhibit FS glass formation and liquid-liquid phase separation. This polymerization process is well-known to arise in liquids of chalcogenide elements with their well-known propensity to have a two-fold coordination because of their two 'lone pair' electrons [54] and other elements such as C under conditions where sp hybridization of the atomic orbitals is prevalent. [55] The literature is very extensive, but we mention some representative references: S [56-59], P [54,60], and Se [61-63] and simulation indicates that this phenomenon also arises in C [55], Te [64-67], Ge [68], Ga [69,70] and other elements. The literature on liquid-liquid has grown rapidly recently and Tanaka has recently provided an up to date review of this field.[71] Although the chemistry of the atomic species can play a large role in the propensity of the atomic species to form polymeric structures upon cooling or at high densities, the formation of dynamic polymeric icosahedral clusters has been seen also in simulations [72] of hard sphere fluids at high particle concentrations so that many-body effects can also give rise to this type of particle self-assembly process due to a purely entropic driving force, even if the individual particles do not possess an intrinsic tendency towards anisotropic bonding at low densities. We further address this equilibrium polymerization perspective of glass-formation in Appendix C, and in our discussion below, since this topic is also peripheral to our main task of investigating the JG $\beta$-relaxation process from a dynamic heterogeneity perspective and we finally mention in this section some nomenclature and basic phenomenology of materials undergoing fragile-strong glass-formation.

Angell [73,74] has suggested an attractive general viewpoint of GF liquids undergoing FS type glass-formation, and we mention his work because his ideas have significantly influenced our approach to describing these materials. In particular, he suggested that this broad and clearly distinct class of GF liquids, subsuming water, silica, $BeF_2$, and some metallic GF materials, is "crystal-like in character while remaining strictly aperiodic", and that this novel state of matter should be intermediate between quasi-crystals and ordinary GF liquids. This is an attractive physical description of what are now called effectively 'hyperuniform' materials. [75,76] He has also introduced a 'bond lattice' model [74,77,78] to rationalize the thermodynamics and dynamics of these unusual GF liquids which have no or very little evidence of a peak in the specific heat as the material goes out of equilibrium, a defining feature of the glass-transition in many experimental studies of 'ordinary' GF materials, while these materials generally exhibit a thermodynamic lambda transition in the middle of the glass-formation temperature range, characterized by a peak in the specific heat resembling a thermodynamic melting transition, as found in anharmonic crystalline materials. Correspondingly, he further suggested that defective 'plastic crystals' exhibiting orientational rather than translational disorder and superionic crystalline materials, and even globular proteins should be included in this same general class of materials. In accord with these remarkable suggestions, we have indeed seen striking similarities between the dynamics of our Al-Sm metallic glass and the dynamics of superionic $UO_2$, a highly anharmonic *crystalline* material, and in our previous simulations of superionic $UO_2$ we observed that the *T* dependence of this material indeed resembles enzyme catalysis to a remarkable degree. [79] Angell's ideas regarding the FS material will be continuously encountered below. We also draw upon similarities between the dynamics and thermodynamics of our simulated Al-Sm metallic glass and the corresponding properties of a particularly well-studied fluid undergoing a FS transition, water. This 'analogy' has been useful to us because it has suggested many properties to investigate to determine if we see the many 'singular' properties observed previously for water. We next turn to the main topic of our paper- diffusion and relaxation processes and quantifying dynamic heterogeneity and cooperative motion and collective excitations in the $Al_{90}Sm_{10}$ metallic GF liquid.

As a final point relating to terminology, we note that superionic transitions in crystalline materials are classified as being Type I and Type II class depending on whether $C_p$ exhibits a discontinuous jump suggestive of a first order phase transition or exhibits rounded shape reminiscent of a second order (lambda) transition. Applying this classification scheme to GF materials exhibiting a FS transition our results below indicate that our Sm-Al metallic GF material exhibits a Type II or 'lambda transition'.

**B. Diffusion and Relaxation Times in an $Al_{90}$ $Sm_{10}$ Metallic Glass**

Standard computational methods are used to determine the diffusion coefficient and structural relaxation time of the atomic species of our model metallic GF liquid. In Figure 1(a), we show an Arrhenius plot of the average self-diffusion coefficient of all atoms obtained from the standard equation, $D = \frac{1}{6} \lim_{t \to \infty} \frac{d<\Delta r^2>}{dt}$, $<\Delta r^2(t)> = \frac{1}{N}\sum_i (\mathbf{r}_i(t) - \mathbf{r}_i(0))^2$, in the limit of long times where $\mathbf{r}_i(0)$ and $\mathbf{r}_i(t)$ are particle's initial and final atom positions after time $t$, respectively, and $N$ is the total number of atoms. This mean diffusion coefficient is defined as the composition weighted (atomic fraction) average of the Al and Sm components, but here we focus on the average $D$. The mean square displacement data for this system is typical for GF liquids and this data both shown in log-log and linear scales is for reference in Supplementary Information (SI).

We define the $T$ dependent activation energy for diffusion from the relation,

$$\Delta G_D(T) \equiv -k_B T \ln(D/D_{oH}) \qquad (1)$$

where $D_{oH}$ is an attempt prefactor related to the vibrational frequency, the geometry, effective jump distance, and activation entropy, [80] and we show $\Delta G_D(T)$ determined from this relation in the inset of Fig. 1(a). The $D$ subscript of $\Delta G_D$ indicates that we are considering the activation free energy for diffusion; other transport properties in general have their own activation free energies and are designated correspondingly in a similar way. It is apparent from this figure that the activation free energy $\Delta G_D(T)$ changes in a sigmoidal fashion with $T$, indicating the transition from a high $T$ Arrhenius regime to another Arrhenius regime at low temperatures. $\Delta G_D(T)$ in these high and low $T$ regimes is designated as $\Delta G_{DH}$ and $\Delta G_{DL}$, respectively. We further report the activation enthalpy $\Delta H_D(T)$ and activation entropy $\Delta S_D(T)$ in the high ($\Delta H_{DH}$, $\Delta S_{DH}$) and low temperature regimes ($\Delta H_{DL}$, $\Delta S_{DL}$) in Table 1, these values being obtained from a fit of our $D$ data to Eq. (1) in the high and low $T$ Arrhenius regimes. In the low $T$ Arrhenius regime, we designate the prefactor as $D_{oL}$ to differentiate it from its high $T$ Arrhenius regime counterpart $D_{oH}$ indicated in Eq. (1). Equation (1) takes the high $T$ Arrhenius regime where transition state theory is assumed to applicable, explaining the explicit occurrence of $D_{oH}$ in this equation.

The *T* dependence of $\Delta G_D(T)$ is the central focus of the string model of glass-formation and below we will compare our estimate of $\Delta G_D(T)$ to the predictions of this model. In particular, the string model predicts that $\Delta G_D(T)$ is proportional to the average length *L* of strings of atoms moving collectively as part of the barrier crossing events associated with molecular diffusion. Of course, this relationship between the 'degree of cooperative motion', i.e., the number of particles moving cooperatively and $\Delta G_D(T)$ is a basic premise of the Adam and Gibbs (AG) model of glass-formation [81] where the extent of cooperative motion is often designated as $z(T)$ so that $\Delta G_D(T)$ formally equals, $\Delta G_D = \Delta G_{DH}\, z(T)$ and *D* is predicted to equal,

$$D(T) = D_A \exp[-\Delta G_{DH}\, z(T) / k_B\, T + \Delta G_{DH} / k_B\, T_A], \quad T < T_A \tag{2a}$$

$$D(T) = D_{oH} \exp(\Delta S_{DH} / k_B) \exp(-\Delta H_{DH} / k_B\, T), \quad T \geq T_A,\; z = 1 \tag{2b}$$

and by mathematical consistency $D(T)$ at the onset temperature $T_A$ for non-Arrhenius diffusion is defined to equal,

$$D_A \equiv D(T_A) = D_{oH} \exp(-\Delta G_{DH} / k_B\, T_A). \tag{2c}$$

The attempt frequency or average collision frequency can be calculated from the initial decay of the velocity autocorrelation function and the determination of this quantity is discussed in Sec. D of the SI of our companion paper focusing on the fast dynamics of the present Al-Sm material [4]; values of the collision relaxation time, the Einstein relaxation time $\tau_E$ or the inverse of the average collision rate, are indicted as a function of *T* in Fig. 3 for the reader's convenience. We see that the average collision frequency, as quantified by $\tau_E$, varies relatively slowly with *T* and, correspondingly, this variation can be expected to have some influence on the diffusion coefficient prefactor, $D_{oH}$. Nonetheless, we follow the usual assumption of taking $D_{oH}$ to be constant with atypical order of magnitude value, as described by Brown and Ashby. [82]

In the low $T$ Arrhenius regime, far below $T_A$ where $z(T)$ again becomes constant $z^*$, we have, $D(T) = D_{oH} \exp[- z^* (\Delta G_{DH}) / k_B T] = D_{oH} \exp[z^* (\Delta S_{DH}) / k_B] \exp[- z^* (\Delta H_{DH}) / k_B T]$, so that the low $T$ diffusion coefficient prefactor becomes, $D_{oL} = D_{oH} \exp[z^* (\Delta S_{DH}) / k_B]$, and the low $T$ 'activation energy' $\Delta H_{DL}$ correspondingly equals, $\Delta H_{DL} = z^* \Delta H_{DH}$. The transition between high and low $T$ Arrhenius regimes has been observed in silicate and other inorganic glasses of geophysical significance and in this context Doremus [83] has defined a measure of 'fragility' appropriate for GF liquids exhibiting FS glass-formation. In particular, his fragility parameter is defined by the ratio, $R_D = \Delta H_H / \Delta H_L$.

Brown and Ashby provide an extensive review of the activation parameters and $D_{oH}$ for crystalline metallic materials [82] and Faupel and coworkers [84] review these transition state theory parameters for metallic GF materials. Together, these extensive reviews cover a wide range of atomic species and summarize numerous correlations and theoretical results that have accumulated for these parameters for metallic materials over many years of intensive studies by many groups. The determination of the characteristic temperatures of glass-formation, such as $T_A$ and the glass transition temperature $T_g$, involves considerations that are specific to our Al-Sm metallic glass material and the estimation of these important temperatures is discussed in SI.

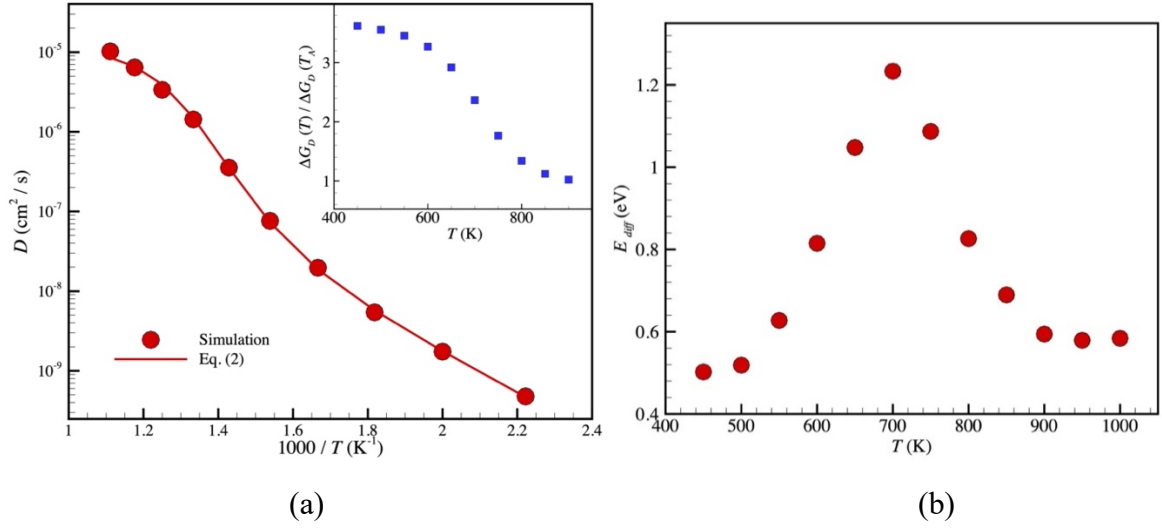

(a)                 (b)

**Figure 1.** Average diffusion coefficient, activation energy and differential activation energy (see text). (a) Arrhenius curves for the average diffusion coefficient $D$ of $Al_{90}Sm_{10}$ metallic glass. $D$ is the composition weighted average of the component atoms, as described in our previous work on $D$ in Cu-Zr metallic glasses. [8,85] The inset shows the activation free energy, normalized by its value at the onset temperature for non-Arrhenius dynamics, $T_A$. (b) Differential activation energy $E_{diff}$ for $D$ for an $Al_{90}Sm_{10}$ metallic glass.

**Table I. Activation Energy Parameters and Pre-exponential Factors of $Al_{90}Sm_{10}$ Metallic Glass in High and Low Temperature Regimes of Glass-formation.**

| $T$ Range | $\Delta H$(eV) | $\Delta S$ (eV/K) | $D_{oH}$ (cm$^2$/s) | $D_{oL}$ (cm$^2$/s) |
|---|---|---|---|---|
| $T < 450$ K | $\Delta H_{DL} = 0.49$ | $\Delta S_{DL} = 1.05 \times 10^{-4}$ | --- | $3.3 \times 10^{-4}$ |
| $T > 920$ K | $\Delta H_{DH} = 0.155$ | $\Delta S_{DH} = 3.32 \times 10^{-5}$ | $4.3 \times 10^{-5}$ | --- |

Many of the standard models and phenomenological expressions describing the relaxation time of GF fluids predict a diverging activation energy upon cooling and all such models can thus be excluded from consideration for our Al-Sm metallic GF material because the activation energy instead plateaus to a constant value in the low $T$ 'glass' state. We also note that the fitted values of $\Delta H$ and $\Delta S$ in Table I in the high and low $T$ Arrhenius regimes both change by a factor of about 3 the high and low $T$ regimes, this activation energy ratio corresponding to $z^*$ and $R_D$ in the discussion above. This change in $\Delta S$ corresponds to a change in the prefactor of $D$ by a factor of 10. In the inset of Fig. 1(a), we have normalized $\Delta G_D(T)$ by its value $\Delta G_{DH}$ at $T_A = 927$ K (See SI for the determination of $T_A$) to emphasize this sigmoidal change in the activation free energy upon cooling. Note that the 'renormalization' of the activation energy parameters in the glass state from values observed in the high T Arrhenius regime implies a kind of entropy-enthalpy compensation [84,86-89] that derives from a change in the degree of cooperative motion in cooled liquids. We then obtain a clear physical interpretation of the Doremus fragility parameter $R_D$ as measure of the change in the extent of cooperative motion between the 'simple fluid' regime where Arrhenius diffusion is exhibited and the fluid is dynamically homogeneous, and the low temperature 'glass regime' where Arrhenius diffusion again is recovered and changes in DH have apparently stabilized, at least in the case of materials exhibiting FS glass-formation.

It is currently not clear whether this low temperature Arrhenius regime exists in 'ordinary' glass-forming liquids, although there have been many observations suggesting this as a possibility. [90,91] It is certainly true that ordinary glass-formers exhibit a stronger tendency to go out of equilibrium upon cooling than FS liquids because the transition to strong glass-formation at low $T$ slows down the rapid increase of the relaxation time in the low $T$ regime in comparison to ordinary GF liquids. This appears to make the entire range of the glass-formation process more discernable in FS GF liquids both by simulation and measurement. In our discussion below and in Appendices A and B, we show that the thermodynamic and dynamic properties of materials exhibiting FS glass-formation exhibit singular features near the $T$ at which the apparent activation energy has a peak in Fig. 1 b. These singular features are also found in 'ordinary' GF liquids, although this 'liquid-liquid' transition may require special care to detect; See Appendix A for a discussion. We are then inclined to believe that the singular features exhibited by materials exhibiting FS glass-formation also occur in all GF liquids, although the strength of these singular features shows a greater or lesser degree of expression in different GF liquids. If this point of view can be established, then FS glass-formation is not really a distinct physical process than 'ordinary' glass-formation.

**C. Quantification of Fragile-to-Strong Glass-Formation**

It is evident that the activation energy, as determined from Eq. (1), does not exhibit the maximum mentioned above as being characteristic of liquids undergoing a FS transition. This singular feature becomes apparent in our data if we adopt a definition of the temperature dependent activation energy often considered by experimentalists in which *local slope* of the Arrhenius plot for *D,* and other transport properties is taken to define the 'apparent activation energy'. Although this local slope is not the true activation energy defined by Eq. (1), except in limits where the activation energy does not vary with *T*, this definition of a temperature dependent effective activation is commonly reported experimentally and then requires some consideration. This definition of apparent activation energy, which we term the 'differential' activation energy $E_{diff}$, is instead a generalization of the definition of the 'steepness' index *m* [92], a well-known measure of the fragility of glass-formation (*m* is just the local slope of the Arrhenius plot at the glass transition temperature $T_g$, normalized by the energy factor $k_B T_g$ to make this quantity dimensionless.)

It does not seem to be generally appreciated that $E_{diff}$ can differ from its definition based on Eq. (1) by *orders of magnitude* [93] and, moreover, their *T* variation can be *qualitatively different.* The deviation between these two definitions of temperature dependent activation energy are especially great in GF liquids undergoing a FS transition. In Figure 1(b), we show the 'differential activation energy' $E_{diff}$ which is estimated from the derivative of the *D* data in Fig. 1(a). We see that this alternative measure of activation energy shows a peak near 700 K. The significance of this characteristic temperature will be discussed at length below in terms of other phenomena that occur near this *T*. It is emphasized that no peak of this kind exists in $E_{diff}$ for ordinary GF liquids.

This general pattern in the *T* dependence of the apparent activation energy in our $Al_{90}Sm_{10}$ metallic glass is observed in fluids exhibiting a FS transition, such as water [94-96] and other GF liquids discussed below. Accordingly, we designate the *T* at which $E_{diff}$ exhibits a maximum (or the true activation energy exhibits an inflection point), the FS transition temperature, $T_{FS}$ (See SI).

The structural relaxation time $\tau_\alpha$ is a basic property quantifying the dynamics of fluids and we may estimate this quantity from the decay of the self-intermediate scattering function $F_s(\mathbf{q}, t)$,

$$F_s(\mathbf{q}, t) = <\exp(-i\mathbf{q}[\mathbf{r}_i(t) - \mathbf{r}_i(0)])> \qquad (3)$$

which is simply the Fourier transform of the particle displacement distribution function $G_s(\mathbf{r}, t)$ from which the mean square displacement above is obtained as the second moment (The Fourier transform variable **q** is often termed the scattering 'wave vector'). This is a standard quantity in liquid state physics because of the observation of this quantity in incoherent neutron scattering measurements, and we show this quantity in Fig. 2(a) for our $Al_{90}Sm_{10}$ metallic GF liquid. This quantity exhibits a first decay on a near *T* independent timescale on the order of a 0.1 ps, followed by transient plateau and a second decay on a timescale that is generally highly *T* dependent.

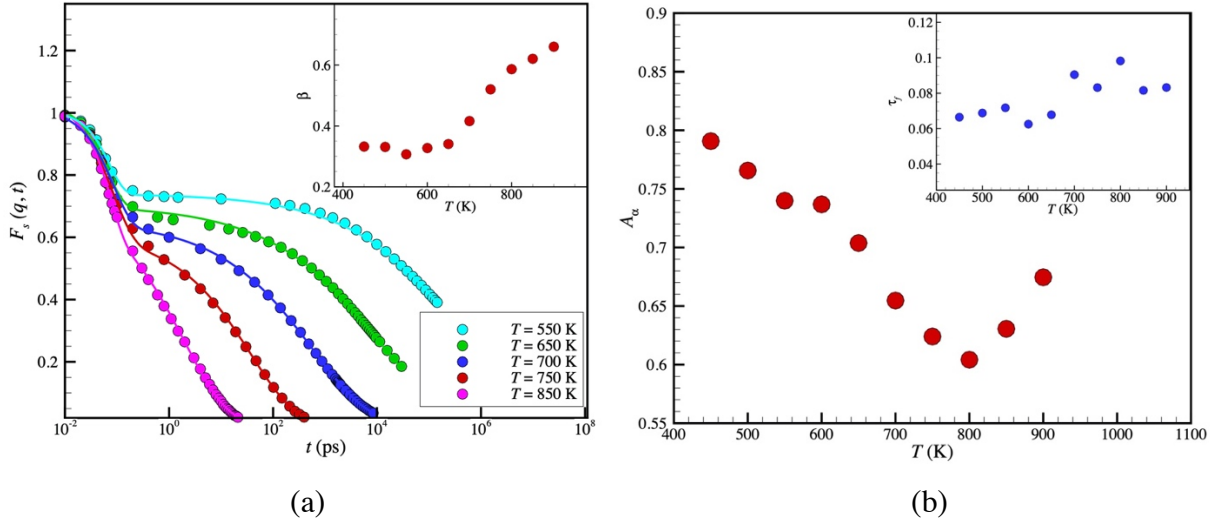

**Figure 2.** Self-Intermediate scattering function and associated parameters. (a) Self-Intermediate scattering function of an $Al_{90}$ $Sm_{10}$ metallic GF liquid for a range of $T$. (b) Temperature of the non-ergodicity parameter and fast relaxation time, $\tau_f$ of an $Al_{90}$ $Sm_{10}$ metallic GF liquid for a range of $T$. Note the value of $\beta$ in the inset of Fig. 2(a) approaches a value near 1/3 at low $T$, which is the value predicted by Douglas and Hubbard for string-like clusters of particles having locally-preferred packing or bonding. [97]

Following previous work [2,98], $F_s(\mathbf{q}, t)$ with $q_0$ corresponding to the interparticle distance is fit by the relation,

$$F_s(q_0, t) = (1 - A_\alpha)e^{-(t/\tau_f)^{\beta_f}} + A_\alpha e^{-(t/\tau_\alpha)^\beta} \qquad (4)$$

where we find the fast beta relaxation decay to be described by a stretched Gaussian ($\beta_f \approx 1.5$) and the $\alpha$ relaxation at long times by a stretched exponential decay where the exponent $\beta$ varies with $T$ (See inset to Fig. 2(a)). The fit of our data to this general form also yields the plateau in $F_s(\mathbf{q},t)$ ('non-ergodicity parameter') $A_\alpha$, the fast relaxation time $\tau_f$ and the '$\alpha$-relaxation time', $\tau_\alpha$. Our estimates of $A_\alpha$ and $\tau_f$ are indicated in Fig. 2(b), where we see a general trend for $A_\alpha$ to become stronger and the fast relaxation to be weaker upon cooling, at least at low $T$. A similar overall trend at low $T$ was observed in recent simulations of a coarse-grained polymer melt. [2] The fast relaxation time $\tau_f$ is on the order of 0.1 ps, a value qualitatively consistent with inelastic x-ray scattering measurements on liquid Na where $\tau_f \approx 0.2$ ps in this range of $q$.[99] We see that $\tau_f$ increases weakly with $T$ (see Figure 3), as observed in recent simulations of water. [100] The increase of $A_\alpha$ starting from a $T$ near the middle of the glass-formation range is not seen in ordinary glasses and we shall see this type of non-monotonic variation in many of the dynamic and thermodynamics properties of our $Al_{90}Sm_{10}$ metallic GF liquid.

The strong and often non-Arrhenius temperature variation of $\tau_\alpha$ is perhaps the defining feature of GF liquids and we show this quantity in Fig. 3, along with $\tau_f$, and other relaxation times discussed below. Given the importance of this property, we show also $\tau_\alpha$ separately for the Al and Sm atomic species. The relaxation time of the relative mass Sm atoms (red triangles) is evidently very slow by comparison to the Al atoms (blue diamonds), where this becomes greatly amplified at low $T$. Apart from the different relaxation times of these different species, it is qualitatively apparent from these plots that multiple components lead to rather distinct relaxation times in the fluid, an effect that probably explains the multiple $\tau_\alpha$ relaxation processes observed in mechanical measurements on this, and other metallic GF fluids. [10] Large mass asymmetries should engender this type of behavior rather generally. It is also apparent that there is a hierarchy of relaxation times that accompanies a hierarchy of relaxation processes, these processes feeding into each other and influencing each other as time progresses.

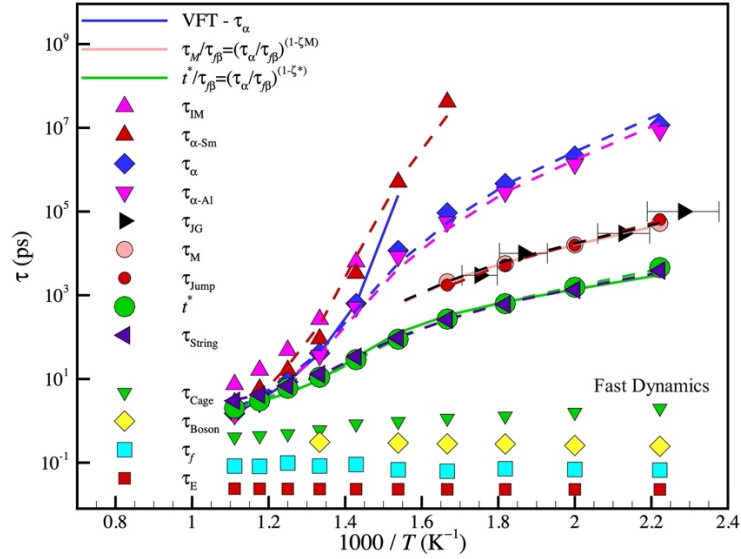

**Figure 3**. Relaxation times and their interrelation. Arrhenius plot of characteristic relaxation times of an Al$_{90}$ Sm$_{10}$ metallic GF liquid for a range of $T$. $\tau_\alpha$ is the structural relaxation time, obtained by fitting self-intermediate scattering function through Eq. (4). $\tau_{\alpha\text{-Al}}$ and $\tau_{\alpha\text{-Sm}}$ are the structural relaxation time for Al and Sm atom, respectively. $\tau_{JG}$ is the Johari-Goldstein slow-$\beta$ relaxation time, determined using simulated dynamical mechanical spectroscopy in a previous study[10]. $\tau_f$ is the 'fast relaxation time' defining the initial decay of the intermediate scattering function. [See Eq. (4)]. $\tau_M$ is the lifetime for mobile particles, determined at the time where mobile cluster size shows maximum [see Figure 10(a)]. $t^*$ is the time where the non-Gaussian parameter shows the maximum [see Figure 6(b)]. $\tau_{IM}$ is the lifetime for immobile particles, determined at the time where the immobile particles size as a function of time shows maximum. [7,11] The 'string time' $\tau_{String}$ is correspondingly defined by the time where the average string length shows maximum [see Figure 14 (b)]. The dashed lines in the figure represent decoupling relations between the average diffusivity $D$ and the corresponding characteristic time, $\tau_i$, i.e., $D / T \sim (1 / \tau_i)^{(1-\zeta_i)}$. We also include estimates of fast dynamics relaxation times determined in our companion work.[4] The Boson peak time $\tau_{Boson}$ is defined as, $\tau_{Boson} \equiv 1/(2\pi\omega_B)$, where $\omega_B$ is the Boson peak frequency. $\tau_{Jump}$ is defined by the inverse of the average jump rate of mobile particle during the lifetime of the mobile particle clusters. The shortest relaxation time $\tau_E$ defines the relaxation time prefactor, which in our metallic glass material has a typical order of magnitude. $\tau_E \sim O(10^{-14})$ s. Specifically, the reciprocal of $\tau_E$ is the mean collision frequency of atoms in potential energy wells defined by surrounding molecules held at their equilibrium positions and is calculated from the decay of the velocity autocorrelation function. [101] $\tau_{Cage}$ is defined as a cage decorrelation time at which the Debye-Waller is $<u^2>$ is defined [4]. The caging time $\tau_{Cage}$ has been identified by some researchers as the 'fast beta relaxation time', although this term perhaps more reasonably applies to $\tau_f$, which describes the shorter timescale of the initial decay of the intermediate scattering function. Relaxation clearly involves a hierarchy of relaxation processes.

The apparent complexity of the relaxation times is mitigated by their common interrelation to $D$ and to each other. In particular, $D / T$ in our cooled metallic GF liquid is commonly observed to exhibit a fractional power-law relation to $\tau$ that is often called the Fractional Stokes-Einstein (FSE) Relation, $D / T \propto (1 / \tau_\alpha)^{1-\zeta}$, and we show in Fig. 4 that this relation linking the rate of diffusion to the rate of structural relaxation holds rather well for our GF liquid. In this sense our GF liquid exhibiting a FS transition is like an ordinary GF liquid. Note that separate powers are observed above and below the λ-transition, where the corresponding exponents are shown in the field of Fig. 4. The more complicated breakdown of Stokes-Einstein relation observed FS glass-formers is shown in the inset of Fig. 4, where the product of diffusivity and the relaxation time is shown to increase at low $T$ (The relation between the peak time in the non-Gaussian parameter and $\tau_\alpha$ also seems more complicated in the present system than a simple power law normally observed in ordinary GF liquids and below we shall see that $D/T$ scales as power law with $t^*$, but the power is not –1, the value normally observed in 'ordinary' GF liquids. [7]) .

A breakdown of the Stokes-Einstein relation naturally arises from the existence of solid-like heterogeneities in the fluid, an effect influences momentum diffusion in the liquid more than mass diffusion. The hydrodynamic 'obstruction model' [102,103] attributes the FSE relation to the presence of finite clusters of immobile particles that persist on times long enough for steady state diffusion to become established. The existence of such long-lived clusters in polymeric and metallic GF liquids of the 'ordinary' variety, [7,8] and the observation of this relation in Fig. 4 suggests to us that such heterogeneities exist in our metallic GF liquid undergoing a FS transition. We shall see below that other relaxation times of our metallic GF liquid are approximately related to $\tau_\alpha$ and to each other through power-law scaling relations, as in ordinary GF liquids. But the change in power-law scaling Fig. 4 suggests that some sort of structural reorganization occurs in FS type GF liquids that is unique or at least prevalent in this class of materials. Apart from subtle quantitative effects, both FS and ordinary GF liquids exhibit a rather similar overall phenomenology in the relation between their relaxation times.

In Appendix B we provide further quantitative evidence that the present material exhibits all the classical identifiable features of FS glass-formation such as singular features in the response functions (specific heat, 4-point density correlation function, etc.) at a common thermodynamic 'λ-transition' that is a signature of these materials. Notably these features arise in a $T$ regime where there is no difficulty equilibrating our material (The structural relaxation times on the order of ns near the λ-transition while our simulation times are performed typically over μs timescales). This material is discussed below because an extended discussion here would distract from our main goal of investigating the JG $\beta$-relaxation process.

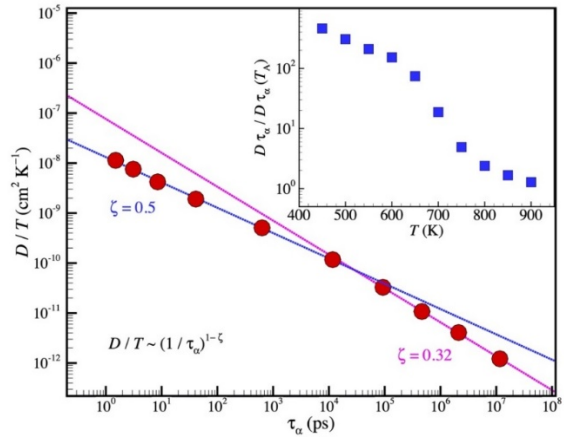

**Figure 4.** Breakdown of Stokes-Einstein relation. We examine the Fractional Stokes-Einstein relation between $D/T$ and $\tau_\alpha$ for an $Al_{90}$ $Sm_{10}$ metallic glass-forming liquid and find that two distinct apparent power law exponents exist in this material in low temperature region ($T < 700$ K) and high $T$ region ($T \geq 750$ K), where the corresponding estimated 'decoupling exponents' are show in the field of the figure. This two-regime behavior has also been observed in water, a prototypical FS glass-former.[104,105] The water measurements required geometrical confinement to suppress crystallization to make the λ-transition observable without the complication of crystallization. The inset shows the breakdown of the Stokes-Einstein relation, where the product of the diffusivity and relaxation time is normalized by their high $T$ values. This data also greatly resembles previous observations on water. [104,106] This two-regime decoupling behavior is not normally apparent in liquids exhibiting 'ordinary' glass-formation, where the anomalies in the thermodynamic response functions at the λ-transition discussed below are also not observed.

**D. Quantification of Standard Measures of Dynamic Heterogeneity**

It is apparent from looking at the distribution function for molecular displacement, the van Hove distribution function $G_s(r,t)$ in our cooled liquid, that the atomic displacement dynamics bears little resemblance to Brownian motion. We illustrate this function for a series of times $t$ for $T = 450$ K in Fig. 5.

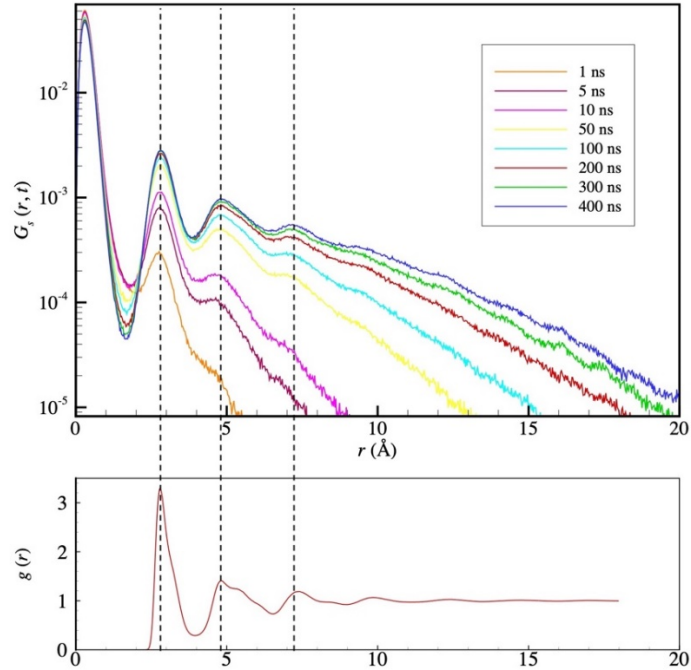

**Figure 5.** The van Hove distribution function $G_s(r,t)$ for atomic displacement for all the particles in an $Al_{90}Sm_{10}$ metallic GF liquid at $T = 450$ K where the pair correlation function is shown directly below with the same radial distance axis indicated. Vertical dotted lines indicate the peaks in $g(r)$, which closely correspond to maxima in $G_s(r,t)$. The sharp maxima beyond the first peak in $G_s(r,t)$ indicate a tendency of atoms to hop to discrete distances that are integral multiples of the interparticle distance.

This tendency toward hopping also becomes more prevalent with a decrease of $T$ and Fig. 6(a) shows $G_s(r,t)$ as function of $T$ at a fixed time, $t = t^*$, where the non-Gaussian parameter shows a maximum [Fig. 6(b)].

The non-Guassian parameter, $\alpha_2(t) = \frac{3<r^4(t)>}{5<r^2(t)>^2} - 1$, provides a conventional and measurable means of estimating the deviation of the displacement dynamics from Brownian motion, for which $G_s(r,t)$ is a simple Gaussian function and $\alpha_2(t)$ is identically zero by definition, shown in Figure 6(b). As with GF liquids generally, $\alpha_2(t)$ peaks at characteristic time $t^*$, which has repeatedly been linked before to the diffusion coefficient $D$ in non-polymeric liquids.[7,8] We check this relation below.

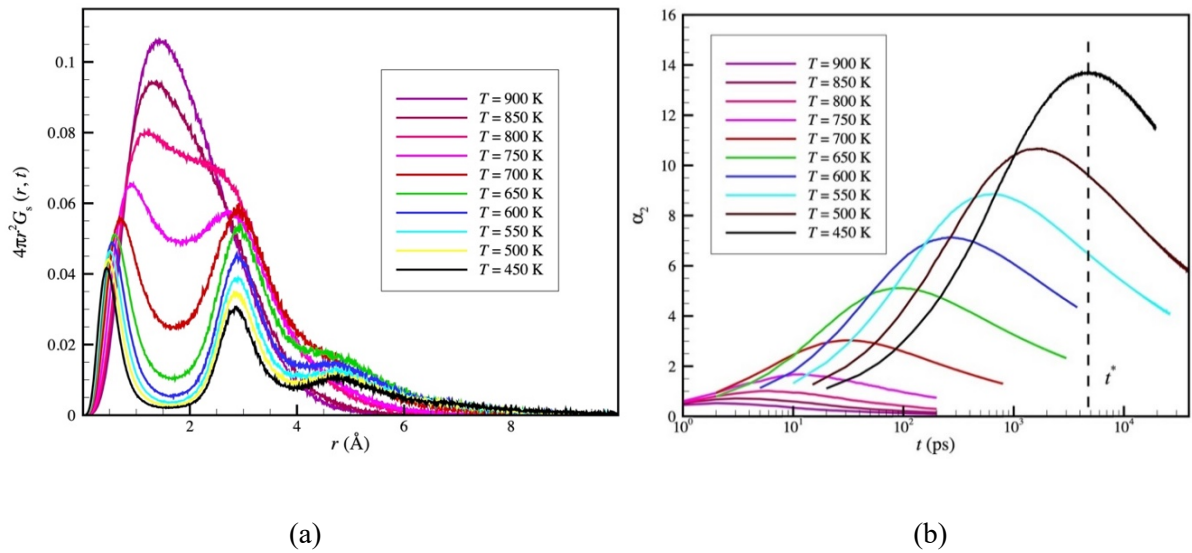

(a)            (b)

**Figure 6.** Van Hove Function and non-Gaussian parameter. (a) The van Hove distribution function $G_s(r,t)$ for atomic diffusion for a $Al_{90}Sm_{10}$ metallic glass-forming liquid for a fixed time $t = t^*$ and a range of $T$. (b) The non-Gaussian parameter $\alpha_2(t)$ for atomic diffusion for an $Al_{90}Sm_{10}$ metallic glass-forming liquid. $\alpha_2(t)$ peaks in the time regime over which the particle displacement is caged (See plot of $<r^2(t)>$ in SI).

Now if we step back and ask ourselves what the atoms actually doing to give rise to this non-Gaussian behavior, then we find that the molecular dynamics is heterogeneous on both spatial and timescale. Figure 7(a) shows the displacement of the atoms at $T = 450$ K over 100 ns period of simulation time where we see that the displacements are highly concentrated on the smaller Al atoms, and that these regions of relatively high mobility are concentrated in band-like regions. The heavier Sm atoms are hardly moving at all and there are also large regions in which the Al atoms are locally tightly packed which are also relatively immobile. This type of dynamics is said to be 'dynamically heterogeneous'.

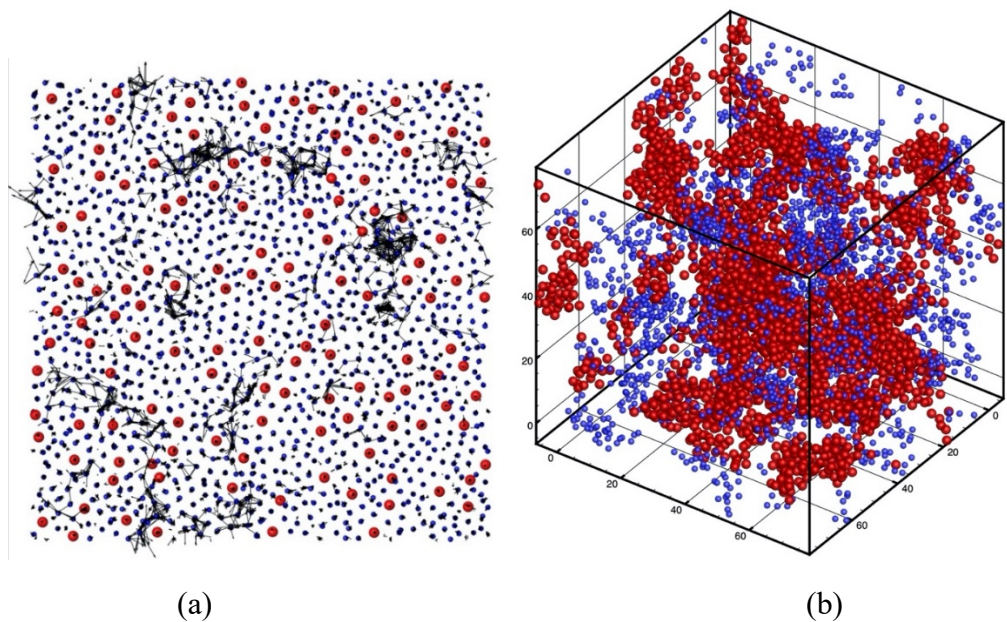

(a)                                               (b)

**Figure 7.** Particle motion visualization. (a) Illustration of particle displacements in a slice of 6 Å thickness $Al_{90}Sm_{10}$ metallic glass-forming liquid at $T = 450$ K and over a timescale $t = 100$ ns (each vector represents a displacement within a time interval $\Delta t = 2$ ns). (b) Typical atomic configuration of mobile (blue) and immobile (red) particles in an $Al_{90}Sm_{10}$ metallic GF liquid at $T = 700$ K.

As illustrated at length previously in ordinary polymer and metallic GF liquids, the highly mobile and immobile particles, relative to the expectations of dynamically homogeneous liquids in which the particles are all undergoing ordinary Brownian motion, are forming dynamic clusters in GF liquids whose average size grows upon cooling and it has previously been demonstrated in simulation that these clusters have direct significance for understanding diffusion and relaxation in GF liquids [7,8], the more mobile particles have a large influence on the rate of diffusion, and the immobile particles greatly influencing the structural relaxation time. The mobile particles are mathematically identified as in previous studies[107,108] $a < | \mathbf{r}_i(t^*) - \mathbf{r}_i(0) | < b$, where constants $a = 1.6$ Å and $b = 4.1$ Å that can be determined from the van Hove correlation function, and the immobile particles are defined as the 5 % least mobile particles in the system within the lifetime of immobile particles. These dynamic clusters arise in our GF liquids as they appear to do for all GF liquids and in Fig. 7(b) we illustrate these mobile and immobile clusters for $T = 700$ K. The nature of these clusters is similar to our past studies, so we are brief in our discussion here.

Although the mobile and immobile particle clusters seem to superficially resemble the clusters observed in previous studies, we see some subtle differences in the geometry of the mobile particle clusters that are notable for this type of metallic GF liquid that might have some relevance to the FS transition. In Fig. 8, we see that the size distribution has the typical power-law distribution with exponential cut-off that we normally see in ordinary GF liquids [7,11], but we see that the distribution progressively approaches an approximately exponential distribution at low $T$. This trend is amplified in Fig. 9(a) where we examine the scaling of radius of gyration ($R_g$) and Fig. 9(b) where we examine the eigenvalues of the radius of gyration tensor. The fractal dimension $d_f$ and shape of the mobile clusters changes from a value associated with branched polymers with screened binary excluded volume interactions or equivalent percolation clusters at high temperatures ($d_f = 2.5$) [109,110] to a fractal dimension and to a shape consistent with self-avoiding or random walk chains. [111] It appears that the cluster topology changes from a highly branched to a nearly linear chain topology, although the mobile particles still exhibit some branching points. Figure 9(b) also reveals that the radius of gyration of the mobile particle clusters exhibits a maximum near $T_\lambda$, which is a phenomenon that certainly does not occur ordinarily in GF liquids. The evolution of the mobile particle cluster distribution with $T$, both in terms of average size, fractal dimension seems to be different in this metallic GF liquid. The generality of this difference is not yet clear, however, nor do we have a compelling explanation of why this change in geometry in the dynamic heterogeneity occurs. We can nonetheless expect this change in geometry to have an impact on the dynamics of the fluid, and we confirm this below.

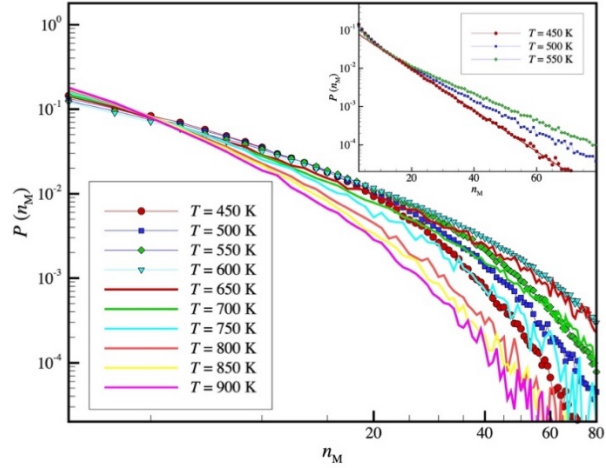

**Figure 8**. Size distribution of mobile particle clusters $P(n_M)$ as a function of $T$ in an $Al_{90}$ $Sm_{10}$ metallic glass-forming liquid.

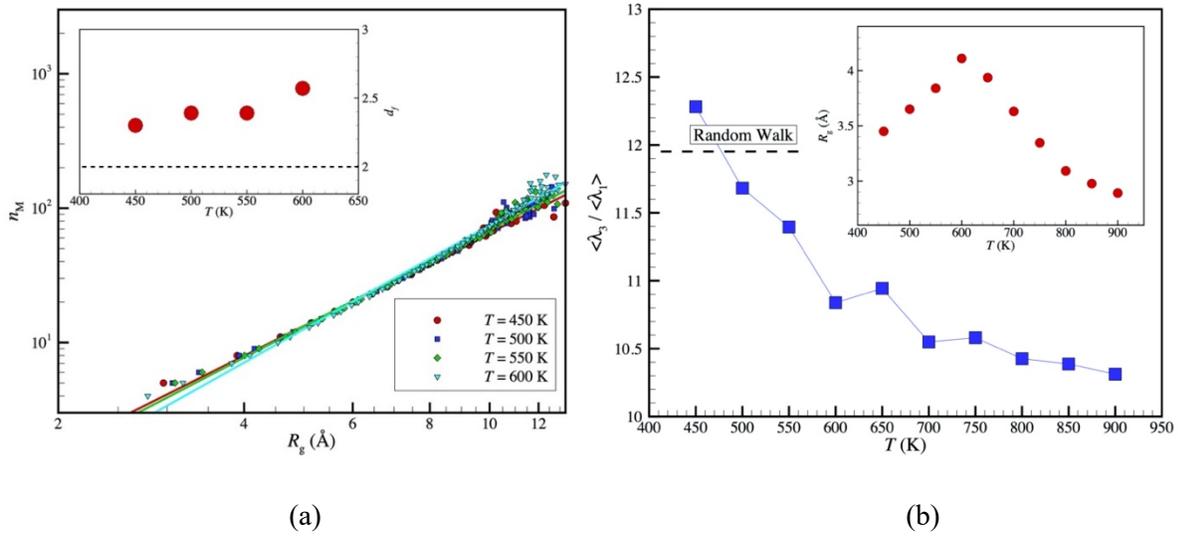

(a)          (b)

**Figure 9.** Fractal properties and average shape of mobile particle clusters. (a) Scaling of mobile particle radius of gyration $R_g$ with its average size $n_M$, $n_M \sim R_g^{d_f}$ where the inset shows $d_f$ as a function of $T$. (b) Radius of gyration $R_g$ and eigenvalue ratio of largest to smallest principal comments of radius of gyration tensor, $<\lambda_3> / <\lambda_1>$, characterizing the shape anisotropy of the mobile particle clusters in an $Al_{90}$ $Sm_{10}$ metallic GF liquid as a function of $T$. The dashed line represents the ratio for random walk [112] and the somewhat larger value at lower $T$ is characteristic of self-avoiding polymers. Notice that the eigenvalue ratio is still somewhat larger at higher $T$ than randomly branched polymers. [113]

As a final point relating to the mobile particle clusters that has great significance for our discussion below is the *lifetime* of the mobile particle clusters. Previous work has shown that the lifetime of the immobile particle clusters scales consistently with the average size (number of particles) in the immobile particle clusters for both polymeric and metallic GF liquids, [7,11] and we find this relation remains to hold for our metallic GF liquid for the limited $T$ range that we can investigate this phenomenon, $T > 700$ K (See SI). In Fig. 10(a), we consider the evolution of the average size of the mobile particle clusters in time for a range of $T$.

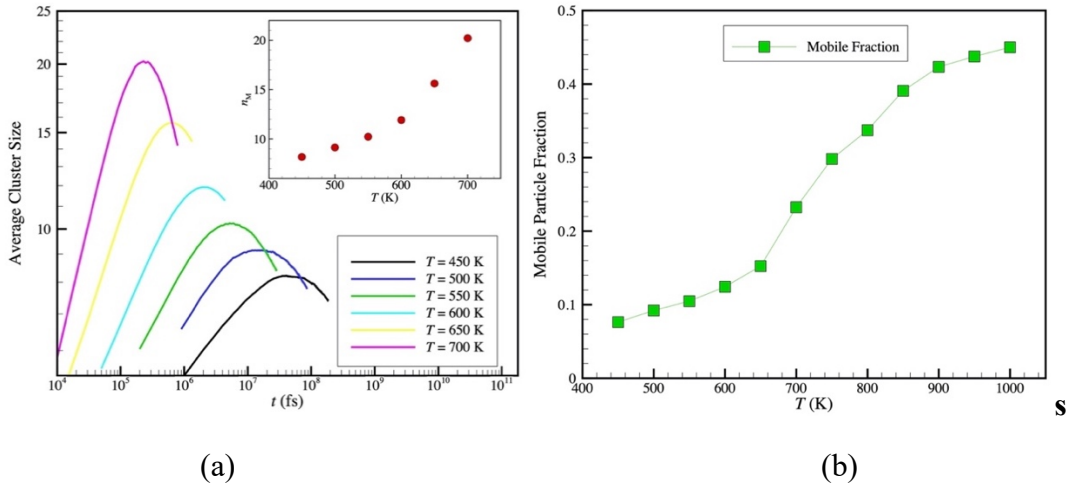

(a)                                                        (b)

**Figure 10.** Size evolution and fraction of mobile particle clusters with $T$. (a) Evolution of the size of the mobile particle clusters in an $Al_{90}Sm_{10}$ metallic glass-forming liquid as a function of $t$ for different $T$. The inset shows that the peak average size grows upon heating, as in the case of heated crystal. [114] (b) Evolution of the average fraction of atoms in the mobile particle state in an $Al_{90}Sm_{10}$ metallic glass-forming liquid as a function of $T$.

A cluster is defined as a group of particles of interest (either mobile or immobile) whose nearest neighbor distance is less than the nearest-neighbor distance based on the radial distribution function. The size of these clusters peaks at a time $\tau_M$, the mobile particle lifetime is defined exactly analogously to the immobile particle lifetime [7,11]. The insert of Fig. 10(a) shows that the average cluster size grows upon heating, as we have seen before for heated crystalline materials.[114] A previous study [7] of the mobile cluster lifetime $\tau_M$ in a model glass-forming polymer melt, studied at relatively elevated temperatures compared to the present work, indicated $\tau_M \approx t^*$, but we find this approximation *no longer holds* in the lower $T$ regime of our Al-Sm metallic GF system where the $\tau_{JG}$ relaxation becomes prevalent. The finding of a separation of the $t^*$ and $\tau_M$ dynamic heterogeneity timescales is an important new finding of the present work. Although we believe this separation of the timescales between $t^*$ and $\tau_M$ is a general result, the separation of timescales needs to be examined in 'ordinary' GF liquids to establish the generality of this observation. Figure 10(b) also shows that the fraction of mobile particles is also progressively increasing upon heating. We shall see that this fraction is quite relevant for the material properties of the metallic glass.

We finally come to point of illustrating the significance of these measure of dynamic heterogeneity in relation to the relaxation dynamics of the material. We see in Fig. 11(a) that the peak time of the non-Gaussian parameter $t^*$ and the peak time for the mobile particle clusters $\tau_M$ both exhibit an approximate power-law relation with the structural relaxation time,

$$(t^* / \tau_{f\beta}) = (\tau_\alpha / \tau_{f\beta})^{1-\zeta^*} \tag{5}$$

$$(\tau_M / \tau_{f\beta}) = (\tau_\alpha / \tau_{f\beta})^{1-\zeta_M} \tag{6}$$

where $\tau_{f\beta}$ is the fast $\beta$-relaxation time defined by the condition that a cage just forms in the liquid and deviations from Arrhenius relaxation occur. The quality of this power-law scaling is not as good in the case of the plot relating $t^*$ and $\tau_\alpha$, echoing the more complicated scaling relation between $D / T$ and $\tau_\alpha$ indicated in Fig. 4 a. There is clearly a change of scaling of $t^*$ with $\tau_\alpha$ upon cooling. However, this change of scaling does not occur exactly at the FS transition, so the significance of this phenomenon is not entirely clear. Previous work [7] has uniformly indicated a clean power-law scaling relation between $t^*$ and $\tau_\alpha$ in ordinary GF liquids. We again have evidence of a subtle change in the pattern of glass-formation in materials exhibiting FS type glass-formation. See Appendices A and B for a discussion of further differences between materials exhibiting FS and 'ordinary' glass-formation.

Time-temperature superposition and power-law relations as in Eqs. (5) and (6) arise when the relaxation process becomes intermittent so that the memory kernel for the relation integral equation becomes a power-law or more generally a homogeneous function. [97,115-117] These exponent values are tabulated in SI, where the small changes in passing through the FS transition are ignored as being a second order effect.

The $\alpha$- and fast $\beta$-relaxation processes *bifurcate* at an onset temperature for complex dynamics $T_A$, which in the present model equals, $\tau_{f\beta}$ = 2.7 ps. In previous work, the 'fast $\beta$-relaxation time' has also been identified with the 'caging time' at which the logarithmic derivative of the mean square particle displacement exhibits a minimum. [7] Here, we note that this onset temperature where $E_{diff}$ starts to deviate from its high constant value may be estimated effectively by the $T$ at which the mean square displacement $<u^2>$ of all the particles at a caging time on the order of ps (defined in the logarithmic derivative just mentioned) equals $<u^2>$ of the mobile particles. All particles become indistinguishable in terms of mobility at this onset temperature. By plotting $<u^2>$ of all the particles and $<u^2>$ for the mobile particles just described we can find the onset temperature for glassy dynamics $T_A$ as the intersection point for these curves as function of $T$, which we illustrate in Fig. 11(b). Evidently, all particles become *indistinguishable* in terms of their mobility at an onset temperature $T_A$ of glassy dynamics, which we may be precisely estimate simply from the $T$ at which $<u^2>$ for all the particles and $<u^2>$ for the mobile particles (based on the definition of mobile particles given above) intersect. We see that this $T$ is around 950 K, consistent with other estimates of $T_A$ (See SI for alternative estimate of TA and other characteristic temperatures of this material).

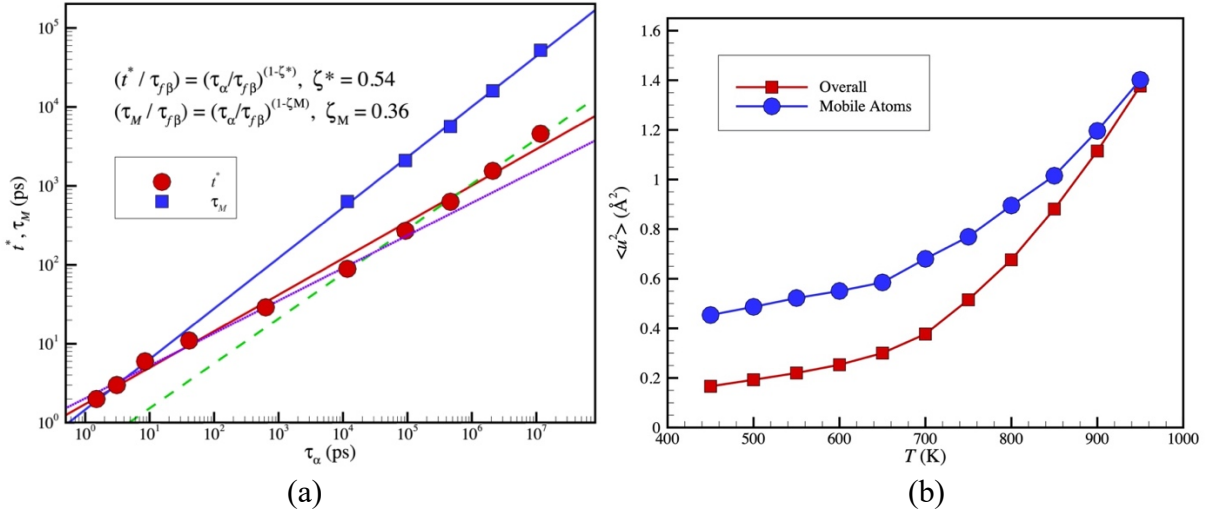

(a)          (b)

**Figure 11.** Relation between non-Gaussian parameter $t^*$ and structural relaxation time and $<u^2>$ as function of $T$. (a) Correlation between peak time of the non-Gaussian parameter $t^*$ and the peak time for the mobile particle clusters $\tau_M$ with the structural relaxation time $\tau_\alpha$ in an $Al_{90} Sm_{10}$ metallic GF liquid. The green dashed line and purple dotted line are the power-law fitting of $t^*$ and $\tau_\alpha$ below and above FS temperature, respectively. (b) Cage size $<u^2>$ of all the particles defined by the mean square particle displacement $<r^2(t)>$ at a 'caging time' $\tau_{cage}$ on the order of ps (defined in the logarithmic derivative just mentioned) for mobile particles and all the particles in the $Al_{90} Sm_{10}$ metallic glass system. All particles in this $Al_{90} Sm_{10}$ metallic GF liquid become *indistinguishable* in terms of their mobility at an onset temperature $T_A$ of glassy dynamics that may be precisely determined as the $T$ at which $<u^2>$ for all the particles and $<u^2>$ for the mobile particle subclass intersect. We see that this $T$ is around 950 K, consistent with other estimates of $T_A$.

Now that we have characterized the structure and lifetimes of the mobile particles, we may ask what significance these structures might have for the dynamics of our GF liquid. In Figure 12, we see that the lifetime of the mobile particles $\tau_M$ overlaps almost perfectly with the Johari-Goldstein relaxation observed in a previous study of our metallic glass fluid. [10] We again find evidence that the mobile particles have a prevalent effect on mass transport in highly cooled liquids. There are multiple types of dynamic heterogeneity and they exert their separate influence on the material dynamics.

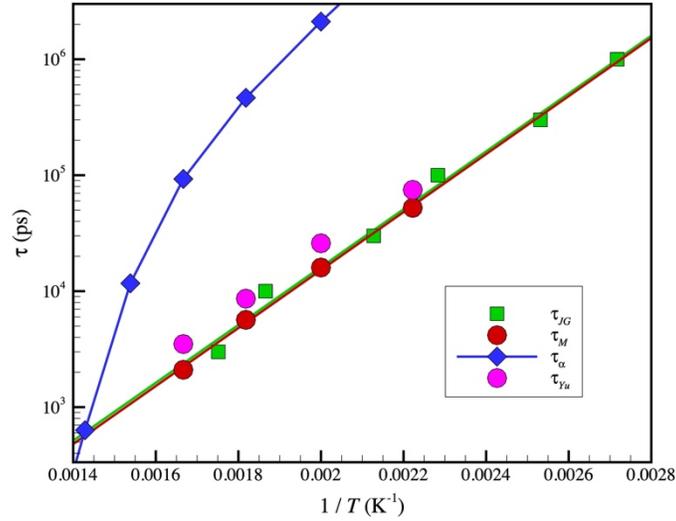

**Figure 12.** Relation between mobile particle lifetime and Johari-Goldstein slow $\beta$-relaxation time and the extrapolated $\alpha$-$\beta$ bifurcation temperature. Comparison $\tau_\alpha$ of an $Al_{90}Sm_{10}$ metallic GF liquid and between the lifetime of the mobile particle clusters $\tau_M$ and Johari-Goldstein slow $\beta$-relaxation time $\tau_{JG}$ determined in a previous study of our metallic glass system. We find evidence that the mobile particles have a prevalent effect on mass transport in highly cooled liquids. Note the extrapolation of $\tau_{JG}$ from a low temperature intersects with $\tau_\alpha$ at $T = 700$ K, which can be defined as the 'bifurcation temperature', $T_{\alpha\beta}$ (See discussion of this characteristic temperature in SI). Apparently, $T_{\alpha\beta}$ is close to $T_c$, but slightly higher than the $T_c$ in the current $Al_{90}Sm_{10}$ metallic GF liquid. This phenomenology is often observed in GF liquids, but it is not universal. [118,119]

We make an additional comment about the recent work of Yu and coworkers [10] who attributed the JG $\beta$-relaxation process to mobile particle clusters that they termed '*strings*'. Relatively high particle mobility does not inherently imply that such particle clusters are involved in collective particle exchange motion, although mobile particles do more prevalent exhibit collective exchange motion than other particles, as we shall investigate in detail in the next section quantifying this type of collective motion in our metallic glass system. We have examined the method of Yu and coworkers for defining 'strings' and found that their method identifies essentially the equivalent of the mobile particle atoms described above. To emphasize this point, we directly compare 'string lifetime' (utilizing the algorithm defined by Yu and coworkers and designated as $\tau_{Yu}$ to avoid confusion with $\tau_{string}$, with our estimate of the mobile particle lifetime) to $\tau_M$ in Fig. 12. Evidently, $\tau_M$ and $\tau_{Yu}$ essentially coincide. We note that other researchers have identified mobile particle clusters based on a method similar to our own, where these clusters have been termed 'strings' [120] or string-like clusters possibly corresponding to the 'cooperatively rearranging regions' of Adam and Gibbs. [121] We emphasize the fractal nature, size distribution, overall spatial size and the temperature dependence of their size and finally the lifetime of mobile particle clusters and strings are normally qualitatively different in cooled liquids so these clusters must be carefully discriminated. The mobile particle clusters are far too large to have any direct relation to the activation energy for transport, and we shall see below that the strings are just the right size to conform to the activation free energy change with temperature, as we shall illustrate below for Al-Sm metallic GF liquid.

We may better appreciate the importance of $\tau_{JG}$ by noting that it is this timescale that directly relates to mass diffusion. We illustrate this in Fig. 13(a) by showing that $D/T$ scales nearly in direct inverse relation to $\tau_{JG}$ (The decoupling exponent is actually - 0.06, which arguably equals 0 to within experimental uncertainty.) We also include the peak time $\tau_{string}$ for the atomic clusters undergoing string-like collective motion described in the next section, where we show these clusters also govern the change of the activation energy shown in the inset of Fig. 17(a).

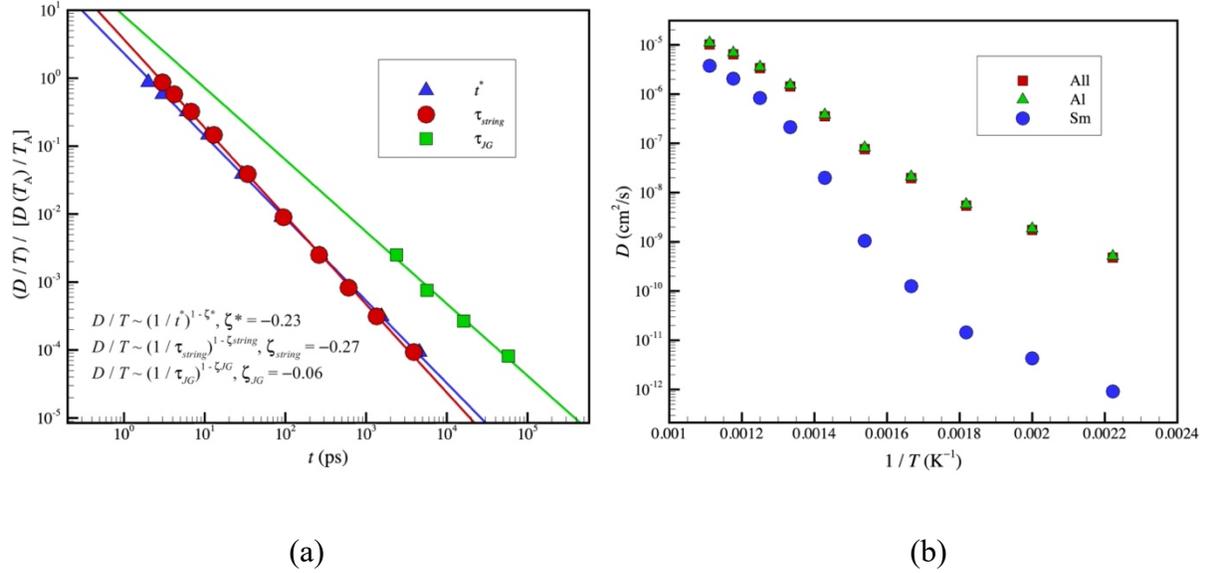

(a)                  (b)

**Figure 13.** (a) Scaling Relation between $D$ and Johari-Goldstein $\beta$-relaxation time. Scaling relationship between $D/T$ and the times: $\tau_{JG}$, $t^*$ and $\tau_{string}$ for $Al_{90}Sm_{10}$ metallic glass-forming liquid. The decoupling exponents are $\zeta_{JG} = -0.06$ so that $D$ correlates very strongly with the inverse of $\tau_{JG}$, but the decoupling exponent takes significantly negative values for $t^*$ and $\tau_{string}$: $\zeta^* = -0.23$ and $\zeta_{string} = -0.27$. We have only previously observed a negative decoupling exponent for superionic $UO_2$. [79] (b) Diffusion coefficients of Al and Sm atoms for an $Al_{90}Sm_{10}$ metallic glass over a range of $T$.

Figure 13(b) shows the diffusion coefficients of the individual atomic species ($D_{Al}$, $D_{Sm}$) in an Arrhenius plot as a function of $1/T$, where we compare to these diffusion coefficients to the average diffusion coefficient $D$ of the entire material. We observe that $D$ practically coincides with $D_{Al}$, and similarly we see from Fig. 3 that the average structural time $\tau_\alpha$ of the entire material roughly equals the structural relaxation time of the smaller metallic glass component, i.e., $\tau_\alpha(Al) \approx \tau_\alpha$. Evidently, the diffusion coefficient of the smaller atomic species (Al) can be orders of magnitude larger than the larger species (Sm), and the average diffusion coefficient and structural relaxation time are both dominated by the fast-moving smaller atomic molecular species. The direct relation between the average $D$ and the JG $\beta$-relaxation time implies from the approximation $D \approx D_{Al}$ that the JG $\beta$-relaxation process should be correspondingly dominated by the rate of diffusion of the smaller atomic species. Measurement studies on model metallic glasses have correspondingly indicated a strong correlation between the diffusivity of the smaller atomic species and the JG $\beta$-relaxation time. [3] The extreme mobility anisotropy in the present material can be naturally understood physically from the rather large difference in the atomic size of Al and Sm. Al as an atomic radius of 1.25 Å, while Sm has an atomic radius 1.85 Å defined in terms of a standard empirical measure of atomic radii [122] so that the Sm atoms are roughly 50% larger than Al atoms, a size anisotropy that we may fairly characterize as being 'unusually large' in magnitude. We also note that Cicerone and coworkers [123,124] have also emphasized a strong correlative relationship between $D$ and the JG $\beta$-relaxation time, which they interpreted in terms of metabasin transitions in the energy landscape of the material.

This discussion of chemically specific aspects of the dynamics of the metallic glass system brings us to consider why the JG $\beta$-relaxation process is so well separated from the $\alpha$-relaxation process in this material. We tentatively suggest that the exceptionally large dynamic asymmetry that accompanies the large atomic size asymmetry is primarily responsible for this large timescale separation in our Al-Sm metallic glass material. We previously found in simulations of Cu-Zr metallic glass materials that a large decoupling means a large separation between the structural relaxation time and the characteristic hopping time associated with particle diffusion. [85] The predominance of the average rate of diffusion in the material by the smaller particles having higher mobility, as noted above, means that smaller particles should give rise to a greater separation between the $\alpha$-relaxation process and the JG $\beta$-relaxation process, provided the JG $\beta$-relaxation process is generally controlled by the rate of atomic diffusion, another phenomenon seen in our simulated Al-Sm material. Zaccone and coworkers have suggested that the time separation between the $\alpha$- and JG $\beta$-relation processes in metallic glasses is primarily due to the mass asymmetry of the atomic species composing the metallic glass [125], an asymmetry that normally accompanies atomic size asymmetry. However, these simple 'geometric' or relative molecular mass interpretations of the timescale separation is probably not completely general. More generally, previous studies of JG $\beta$-relaxation in other materials have suggested that the strength of molecular cohesion of particles with those in their surroundings [126,127] can also be important, in addition to the relative molecular size to the molecules or polymer segments of the fluid to which the molecules are added (Additional discussion on this topic can be found in SI.)

The study of the molecular dynamics processes responsible for the disintegration of the mobile particle clusters requires a detailed of the dynamics of the GF material at times much shorter than $\tau_{JG}$. We investigate this sub-dynamics of the JG $\beta$-relaxation process in a companion paper [4], focusing on 'fast' dynamical processes. In this work, we observe that atomic diffusion in the low $T$ regime in which the JG $\beta$-relaxation is prevalent occurs through sharply defined and large scale 'jump' events at intermittent times, these jumps being mediated by relatively large-scale and *reversible* string-like collective motion on a ps timescale, a collective phenomenon interpreted as arising from localized stable modes and found to be directly related to the Boson peak. We also quantify in this companion work the rate of atomic jump motions driven by these collective localized modes and show the inverse average rate of hopping events *coincides* with the JG $\beta$-relaxation time to within numerical uncertainty, providing some insight at atomistic level of the relation between $D$ and the JG $\beta$-relaxation time just noted. The intermittency of this jump process also has significant ramifications for understanding colored noise and power law relaxation in association with the JG $\beta$-relaxation process. [4] The 'fast' dynamical processes underlying the JG $\beta$-relaxation process are evidently rather rich. We also discuss in our companion paper the compatibility of our simulation observations indicating a direct relation between the lifetime of large clusters of mobile particles and the JG $\beta$-relaxation process and Johari's 'islands of mobility' picture of the JG $\beta$-relaxation process. [128] Our findings are quite consistent with Johari's qualiative interpretaion of the JG $\beta$-relaxation process and can now make quantitative pictures of these formerly hypothetical structures and we may also study the evolution of these dynamical structures in association with aging. See companion paper [4] for visualizations of the dynamic heterogeneity.

**D. Quantification of Cooperative Atomic Exchange Motion – 'Strings'**

To quantify collective particle motion, we must identify the 'mobile' atoms in our system. Since "mobile" atoms are essentially those particles moving a distance $r(t)$ larger than the typical amplitude of an atomic vibration after a decorrelation time $\Delta t$, but smaller than the second nearest-neighbor atomic distance (as suggested by the multiple peaks of the self-part van Hove correlation function shown in Figure 6(a)). The further quantification of collective atom motion requires a consideration of the relative displacement of the particles. Collective atomic motion implies that the spatial relation between the atoms is preserved to some degree as the atoms move. In particular, the reference mobile atoms $i$ and $j$ are considered to be within a collective atom displacement string if they remain in each other's neighborhood, and we specify this proximity relationship by, min $[|\mathbf{r}_i(t^*) - \mathbf{r}_j(0)|, |\mathbf{r}_i(0) - \mathbf{r}_j(t^*)|] < 1.1$ Å. Illustration of typical string-like collective motions is shown in Figure 14(a) and weight average string length as a function of simulation time at different temperatures is shown in Figure 14(b). Atomistic simulations of GF liquids also indicate that the distribution of string lengths $P(n)$ is approximately an exponential function of the number of atoms in the string $n$, i.e., $P(n) \sim \exp(-n/<n>)$, where $P(n)$ is the probability of finding a string of length $n$ at the characteristic time, $t^*$ and $<n>$ is the average string length, $L$. The size distributions of string length at three representative $T$ are shown in Figure 15.

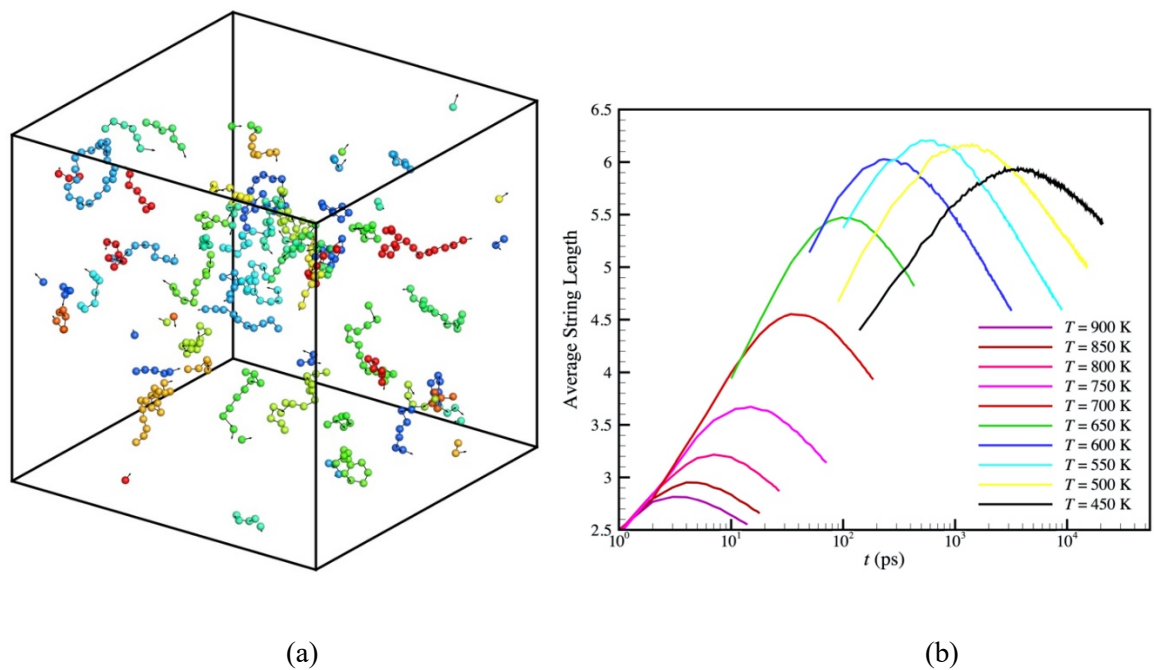

(a)            (b)

**Figure 14.** Illustration of strings in our $Al_{90} Sm_{10}$ metallic GF liquid. (a) Snapshot of the strings identified in our system for $T = 450$ K. (b) Evolution of the average number of particles involved in cooperative exchange motion, the string length $L$, with time, $t$.

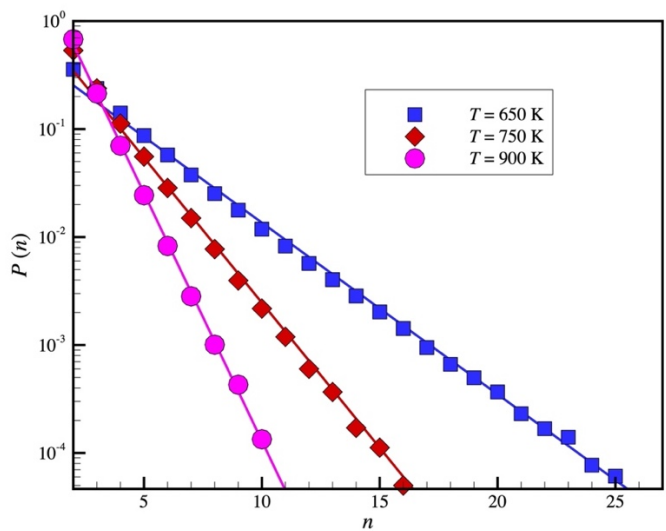

**Figure 15.** Size distribution of strings (See Fig. 14) in our $Al_{90} Sm_{10}$ metallic GF liquid for a range of $T$.

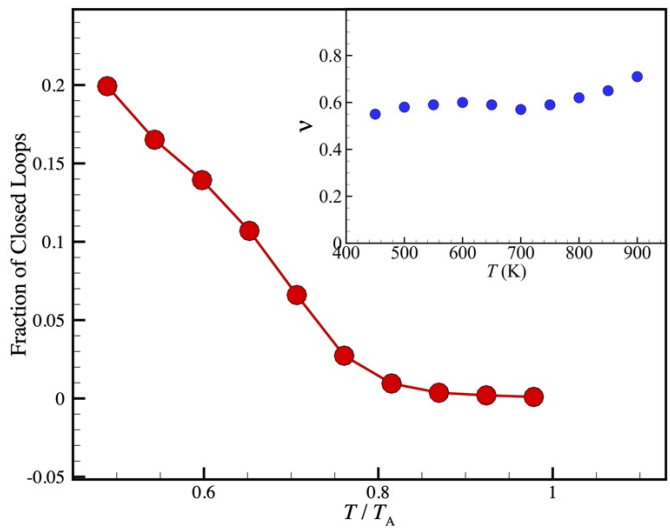

**Figure 16.** Fraction of strings taking the form of closed loops ('ring polymers') as a function of reduced temperature, $T / T_A$. In previous work on the homogeneous melting of crystalline Ni, we also saw a topological transition in which the closed rings at low $T$ break to form linear chains at elevated $T$. Scaling between string radius of gyration $R_g$ and its length $L$ follows $R_g \sim L^\nu$, where exponent $\nu = 1 / d_f$ is shown in the inset at different $T$. The string fractal dimension is initially near $d_f = 5/3$ and progressively approaches the screened excluded volume interaction value of 2 as $T$ is lowed towards $T_g$ where the strings apparently more strongly interact. $\nu$ values close to those for self-avoiding walks and a near exponential length distribution have been observed in previous observations on metallic glass and polymeric glass-forming liquids exhibiting 'ordinary' glass-formation. [7,8] Interestingly, the fractal geometry and size distribution of strings observed in superheated crystalline Ni [114] and in superionic crystalline $UO_2$ [79] are again very similar to our present observations on AL-Sm alloy exhibiting fragile-to strong type glass-formation. The geometry of the strings seems to be quite universal and *distinct* from the mobile particle clusters, which exhibit the universal scaling characteristics of *branched* equilibrium polymers rather than linear polymers.

As in our previous study of collective motion in connection with the homogeneous melting of crystalline Ni[114], the strings undergo a *topological transition* as $T$ is varied. We also saw a topological transition in which the closed rings at low $T$ break open at elevated $T$ (see Figure 16), which apparently played a role in destabilizing the crystal. i.e., 'melting'.

We now come to our test of whether the string model can account for the change of activation energy shown in Fig. 1(a). The string model is formulated as generalized transition state theory in which the activation free energy events associated with diffusion are described by a reaction process involving a catenation of activation events. [1,89,129] Betancourt et al. [1,16]

extended the string model in an important way by considering the collective activation barrier crossing process to be an initiated reaction process, which makes the fragility of glass formation variable and has the consequence of leading to residual collective motion above $T_A$ and a limited extent of collective motion in the low temperature glass state, as in systems exhibiting FS glass formation. Xu et al. [130] provide an extensive review of the string model and related entropy-based theories of glass-formation.

The string model has been discussed previously and here we just note the prediction of this model for the temperature dependence of $D$ in a cooled liquid exhibiting a temperature dependent activation energy based on this equilibrium polymerization reaction mechanism, $D(T) = D_A \exp[-\Delta G_{DH} (L / L_A) / k_B T + \Delta G_{DH} / k_B T_A]$, where $D_A$ is defined by Eq. 2c above. The ratio $L / L_A$ in this equation indicates the relative change of the scale of string-like collective motion below $T_A$, corresponding to the reduced activation free energy, $\Delta G_D(T) / \Delta G_{DH}$, in the string model of glass-formation. [1,89] It is evident also that the 'extent cooperative motion parameter' $z(T)$ of the AG model directly corresponds to the string model parameter, $L / L_A$. The internal consistency of the string model has been validated by the independent calculation of $\Delta G_\tau(T)/ \Delta G_{\tau H}$ (Note subscript $\tau$ denotes the segmental relaxation time is the property whose activation free energy is being considered.) [16,131] for model GF polymer melts. It is noted that the same 'cooperativity factor', $L / L_A$, describes the change in activation free energy for all relaxation times in which a power-law relation holds between $D / T$ and the corresponding relaxation time. The string model has also performed well in the description of pure bulk and thin polymer films, thin films supported on boundaries, polymer nanocomposites, polymers under applied pressure, star polymers, etc. [7,16,89,132-134] and a Cu-Zr metallic GF liquid. [8]

We see that there is rather good accord between $\Delta G_D(T)/\Delta G_{DH}$ in Fig. 17(a) and our independent estimate of $L/L_A$ for the Al-Sm metallic glass material. This is the first time that the string model has been tested for consistency in a metallic material. The is a notable small deviation from $G_D(T)/\Delta G_{DH}$ from $L/L_A$ below $T \approx 500$ K, which suggests to us that non-equilibrium effects are beginning to become at these very low $T$. Previous work on relaxation in thin films indicates that the relation between $L$ and the activation free energy is completely lost when the material goes far out of equilibrium [135] so we believe our results have some qualitative validity in this low temperature regime. Nonetheless, simulation estimates at such low $T$ should be viewed with caution.

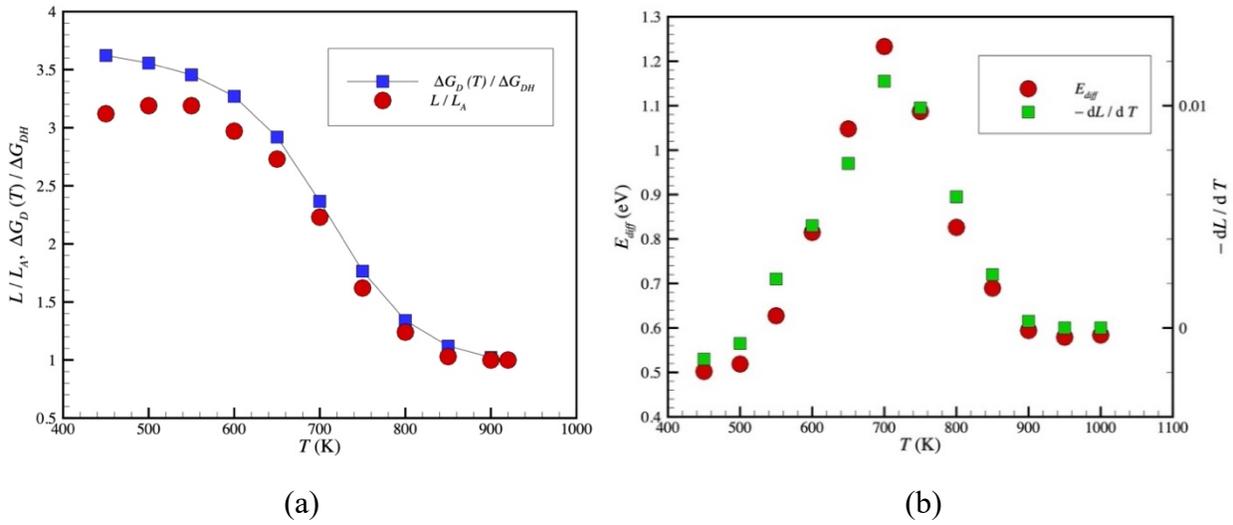

(a)        (b)

**Figure 17.** Normalized activation energy, string length and apparent activation energy. (a) The activation energy for $D$, normalized by its value in the high temperature regime, $T > 950$ K shown in the inset of Fig. 1a is compared to the *independently determined* string length, $L$. The string model can evidently account for the $T$ dependence of the activation, consistent with the view that the strings correspond to many particle transition state events associated with particle diffusion. (b) The differential activation energy $E_{diff}(T)$ obtained from the slope of the Arrhenius plot of or the relaxation time is compared to the differential change in the degree of collective motion, $dL/dT$. The differential activation energy $E_{diff}(T)$ then provides information about the differential change in the degree of collective motion rather than the magnitude of collective motion.

We may also gain insight into the into the $T$ dependence of the differential activation energy, $E_{diff}(T)$, in relation to collective motion in our fluid. In Fig. 17(b), we compare $E_{diff}(T)$ to the *derivative* of the string length, $dL/dT$, where we see that these quantities essentially coincide up to constant of proportionality. Even though the widely reported $E_{diff}(T)$ has no real connection to the real activation energy, this quantity apparently can have direct information about the differential change in the degree of collective motion.

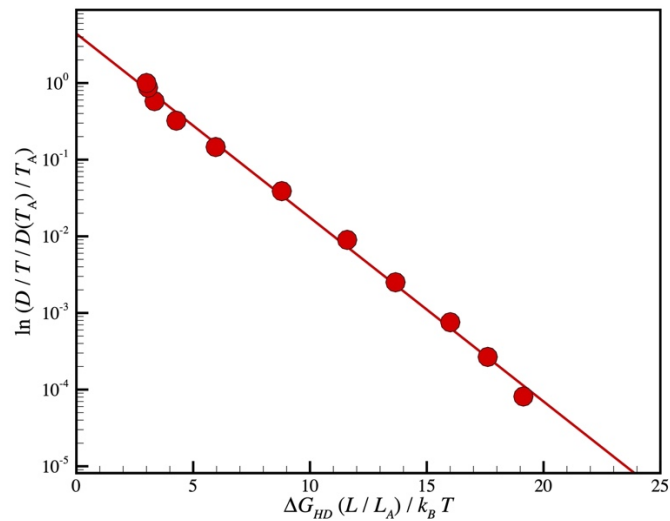

**Figure 18.** Test of the String Model for $D$ data. Comparison of the string model to $D/T$ normalized by $D(T_A)/T_A$ versus $\Delta G_{HD}(L/L_A)/k_B T$, where $\Delta G_{HD} = \Delta H_{HD} - T\Delta S_{HD}$, where $L_A \equiv L(T=T_A)$.

We note that it is not possible to describe the change of the diffusion coefficient from a change of the activation energy alone since we must also account for the change of the activation entropy $\Delta S_{DH}$ in the rate process as well (see Table 1). As noted above, the value of $\Delta S$ in the high and low $T$ regimes of glass-formation are significantly different, which means that the prefactor in the Arrhenius regime has been significantly altered from its value in the low $T$ regime. In the string model, $L/L_A$, plays the role of $z(T)$ in the AG model, and this quantity multiplies both $\Delta H_{DH}$ and $\Delta S_{DH}$. The string model then also predicts this change in the prefactor, even for intermediate $T$ regime where the activation free energy is $T$ dependent. Note that one difference between the string and the AG models of the dynamics is that the collective motion does not disappear in the Arrhenius regime, but the extent of collective motion becomes nearly constant in magnitude, i.e., there is some residual collective motion in the Arrhenius regime so that $L_A > 1$. A typical value of $L_A$ in glass-forming liquids lies in the range 1.2 to 1.4.

## IV.    Conclusions

We investigate the JG $\beta$-relaxation process, along with its relation to the average diffusion coefficient and the $\alpha$-relaxation processes in a simulated Al-Sm glass-forming material exhibiting a fragile to strong transition. Our overall goal was to obtain a clear understanding dynamical structural origin of the JG $\beta$-relaxation process in terms of conventionally defined measures of dynamic heterogeneity. Direct computation shows that the lifetime of the mobile particle clusters coincides with independently estimated estimates of the JG $\beta$-relaxation time from experiments and simulations of mechanical properties of the same material determined in previous work by Sun et al. [10] These findings clearly indicate a well-defined interpretation of the JG $\beta$-relaxation process from a DH perspective, although we still need to confirm this DH interpretation of the Johari-Goldstein relaxation process in 'ordinary' glass-forming liquids at corresponding low temperatures where this relaxation process becomes discernable. The findings of the present work complement observations of a direct relationship between the now standard measure of immobile cluster lifetime and the average structural relaxation time obtained from the intermediate scattering function. We have thus obtained a unified potentially description of the main relaxation processes of glass-forming liquids within picture of DH framework (subject to verification of our findings in 'ordinary' GF liquids). In the end, as in many cases where a phenomenon can be understood with the help of hindsight, our findings are almost 'obvious'. Domains of mobile particles dominate relaxational processes, such as mass diffusion, and domains of immobile particles dominate the rate of structural relaxation. Decoupling derives from the disparity between the lifetimes of these clusters defining DH in the material. We now have well-defined structural entities, albeit whose form being defined by particle dynamics rather than static structure, to study in relation to understand physical aging, the non-linear mechanical properties, shear banding, impact resistance and many other aspects of materials in their 'glass' state where may further expect the DH to evolve in time. [7] Figure 19 illustrate the typical geometrical forms (i.e., mobile clusters and strings) of dynamic heterogeneities in our Al-Sm metallic glass system. The geometrical form of mobile and immobile clusters was illustrated in Figure 7(b).

Although our primary goal was to elucidate general aspects of JG $\beta$-relaxation that might apply to all glass-forming liquids, we found some aspects of the JG $\beta$-relaxation process that are of specific interest in metallic glass-forming liquids and probably other glass-forming liquids derived from liquid mixtures. We observe that the compositionally averaged diffusion coefficient $D$ of all the atomic species is practically identical to the diffusion coefficient of the smallest atomic species (Al), and that the rate of particle hopping of the atoms is directly related to the JG $\beta$-relaxation time, $\tau_{JG}$. Both of these inter-relationships have been suggested in recent experimental studies on glass-forming liquids. [3,123,124]

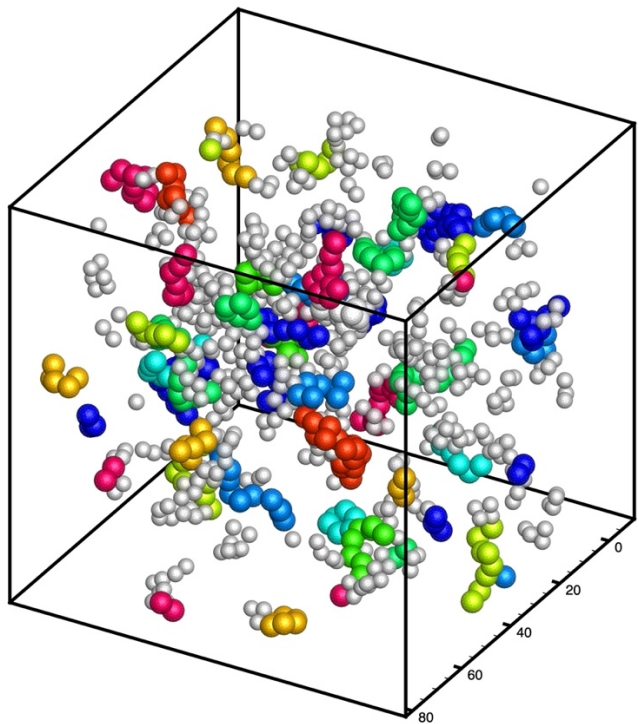

**Figure 19.** Snapshot image showing mobile particle clusters and strings. The colored atoms represent those participating in the string-like collective motion where the different colors indicate different strings (only strings with length longer than 6 atoms are shown for clarity). The grey atoms are the clusters of the mobile particles; these clusters are more branched in nature, as quantified in the text, and these clusters have a longer lifetime than the strings at low temperatures that is related to the Johari-Goldstein relaxation time. In particular, the lifetime of the mobile particle clusters is $\approx 50$ ns at $T = 450$ K, but the lifetime of the strings is a much smaller value, $\approx 3.9$ ns. The clusters of atoms exhibiting string-like collective motion are evidently embedded in the mobile atom clusters. The strings are the particle that move collectively over one representative string lifetime over a much longer time period over which the mobile particle cluster exists. Note this type of geometrical form of DH configurations is also typical for ordinary GF liquids, both in metallic glass and polymeric systems. [7,8] There are also stringlet clusters (not shown) involving highly coherent string-like collective motions on a ps timescale. As discussed in our companion paper focusing on the fast dynamics of the present material, these 'fast' collective atomic motions play a significant role in the disintegration of the mobile particle clusters. We do not show these clusters because they are normally much shorter than the strings and because they occur seemingly everywhere within the mobile particle clusters.

As a by-product of our study, we encountered the phenomenon of FS glass-formation for the first time, providing an opportunity to test the generality of the dynamic heterogeneity phenomena in this important and largely unexplored family of GF liquids from computational/dynamic heterogeneity perspective. This fragile-to-strong transition is reflected both in structural relaxation time and diffusion coefficient, leading to an approximate power-law decoupling relation relating these properties, $D/T \sim (1/\tau_\alpha)^{1-\zeta}$, where $\zeta$ is a material dependent 'decoupling exponent'. An approximate decoupling relation also relate $D/T$ to the peak time in the non-Gaussian parameter $t^*$ where $\zeta$ is negative, $\zeta^* = -0.23$. As discussed above for other properties, the approximate power-law between $D/T$ and $t^*$ with a single exponent is not as compelling as found for ordinary GF liquids, and again we find slightly different effective powers above and below the FS transition. However, we find a much more convincing power-law scaling relation between $D/T$ and the JG relaxation time, $\tau_{JG}$. Moreover, we find the decoupling exponent between these properties $\zeta_{JG}$ is nearly equal 0, supporting a direct relation between $\tau_{JG}$ and $D$.

To relate these dynamical properties to dynamic heterogeneity and cooperative motion, we quantified the lifetime of both the immobile and mobile clusters in our liquid and compared these to the independently determined relaxation times of our system. Finally, we validated the string model of glass-formation, which predicts the activation free energy for $D$ varies proportionately to the average length of string-like cooperative exchange motions in the fluid. We find that the lifetime of the mobile particle clusters coincides with the JG $\beta$-relaxation time, a relaxation process that becomes a prevalent mode of material relaxation at $T$ below the glass transition temperature where the $\alpha$ relaxation process becomes inactive because of the large value of $\tau_\alpha$. Structural relaxation in cooled liquids clearly involves a hierarchy of relaxation processes acting on rather different time and spatial scales.

We have interpreted glass-formation in our metallic glass materials as an equilibrium polymerization process involving the local organization of the Al 'solvent' around the Sm 'solute' particles and the organization of the atomic clusters into larger scale structures having a string-like geometry that defines medium range order in the material. These well-packed regions play a role analogous to grain boundaries in polycrystalline materials and the grain-like bundles of well-packed atoms is surrounded by grain boundary regions having relatively poor structural organization in which atomic motion takes the form of string-like collective exchange motion. This picture is just as we have observed the structure in previous simulation studies of polymeric and metallic glass-forming liquids. [7,8] Polymerization is a kind of ordering process that leads to a drop in the configurational entropy of the material. But the configurational entropy does not drop to zero in the low temperature glass state as in crystals at low temperature. The residual entropy apparently derives from the large conformational entropy of the polymer chains. [136]

There are different general classes of dynamic heterogeneity that are necessary for understanding diffusion and relaxation in glass-forming liquids, relaxation processes related to mass and momentum diffusion, respectively. The mobile particle clusters dominate mobility (diffusion) and the immobile particles dominate the shear viscosity and structural relaxation time. The mobile particle clusters are composed of more primitive particle clusters in which the particles move in a characteristic cooperative exchange fashion ('strings'), as viewed on relatively long timescales comparable to the lifetime of these dynamic clusters. The strings are linear polymeric structures, while the mobile particles have the geometrical characteristics of branched equilibrium polymers. Strikingly, the nature of the dynamic heterogeneity has been found to be the same in simulations of coarse-grained polymer melts, thin supported polymer films, polymer nanocomposites, a wide range of polymer topologies. We have found essentially the same geometrical form of dynamic heterogeneity in Cu-Zr metallic glasses, which exhibit ordinary glass-formation, before our current simulations of the Al-Sm system. The geometrical nature of dynamic heterogeneity in GF liquids appears to be remarkably universal, although the $T$ dependence of this dynamic heterogeneity can differ between these different classes of GF liquids. In addition to the subtle 'kinks' in the decoupling curves relating $D / T$ and the non-Gaussian parameter time $t^*$ and to the structural relaxation time $\tau_\alpha$ in the system exhibiting FS glass-formation, we find that thermodynamic response functions, such as the peak height of the 4-point density function, a common measure of DH, exhibit *maxima* in the vicinity of the fragile-to strong transition. Extrema in the response functions are very characteristic of systems undergoing fragile-to-strong transition, while such features are either too small or absent in ordinary GF fluids. We discuss these singular thermodynamic features for our material in Appendices A and B. There are clearly some aspects of glass-formation in systems undergoing FS glass-formation that are quite distinct from ordinary GF liquids. In the main text of our paper, we have emphasized the common geometrical aspects of the dynamic heterogeneity in these distinct classes of GF liquids. We briefly summarize these general properties of dynamic heterogeneities.

The immobile particles, as in the case of the mobile particles, both resemble branched equilibrium polymers, which adopt the fractal dimension near 2.5, the fractal dimension of randomly branched polymers with screened binary excluded volume interactions, i.e., percolation clusters. We have discussed this fractal geometry aspect of dynamic heterogeneity in our previous works [7,8] where we characterized the dynamic heterogeneity in both polymeric and metallic glass-forming materials to gauge the generality of the phenomenon. In the present work, we have found a new phenomenon in which the mobile particle clusters appear to be adopting a somewhat lower fractal dimension and a shape more reminiscent of random walk polymers, i.e. $d_f = 2$, at the unprecedentedly low temperatures that we investigate from a dynamic heterogeneity perspective. This trend is extremely interesting because this type of behavior has been often observed in simulations of phase transitions, such as in the vortex proliferation model of the superfluid-normal transition and type-II superconductor-normal conductor transition, where there is a first proliferation of polymeric vortex excitations exhibiting approximately self-avoiding walk geometry [137-140], followed by a loop condensation disordered regime in which finite loops spanning the whole material system coexist with finite loop-like strings. [141-143]

In our companion paper, [4] we focus our attention on the fast dynamics of the Al-Sm metallic glass material and the interrelation to the long-time structural relaxation time, Johari-Goldstein process and diffusion processes investigated in the present paper. In particular, we investigate the fast $\beta$-relaxation and the atomic dynamics underlying the Johari-Goldstein $\beta$-relaxation processes, along with the elastic scattering response of GF liquids and the Boson peak in a companion paper. String-like particle motion, having both an irreversible and reversible nature ('stringlets'), occurs in the 'fast' dynamics regime, corresponding to a ps order of magnitude timescale. String-like collective motion associated with localized *unstable* modes facilitate irreversible and intermittent particle 'jumping' events at long times associated with the Johari-Goldstein (JG) relaxation process, while stringlet motion associated with localized stable modes, and corresponding perfectly reversible atomic motion, gives rise to the Boson peak in our system. In particular, we calculate the density of states for both the stringlet and "normal" particles and find that the stringlet particles give rise to a Boson peak while the normal atoms do not. The growth of stringlets upon *heating* ultimately also leads to the "softening" of these excitations, and the Boson peak frequency and shear modulus drop in concert with this softening. The growth of string-like collective motion upon *heating* in the 'fast' dynamics regime is further shown to be responsible for the growth in the intensity of the fast relaxation process.

**Appendix A: Observations on the Nature of the Fragile-Strong Transition**

Recently, it has become appreciated that glass-formation in many materials can follow a strikingly different phenomenology than 'ordinary' glass-forming liquids, which are characterized by apparent activation energies $E_{app}$ for their mass diffusion coefficients, structural relaxation times and viscosities in conventional Arrhenius plots that monotonically increase upon cooling, a pattern of relaxation that many models of glass-formation have sought to rationalize. In some liquids, however, such as water and silica, and also many metallic GF materials as we will discuss below, $E_{app}$ first increases rather sharply upon cooling, but then suddenly drops, and then $E_{app}$ apparently becomes constant again in the glass state, taking a value much lower than its peak value at intermediate temperatures. Because of the change from a rapidly increasing $E_{app}$ to a condition for which $E_{app}$ is nearly constant, this type of GF liquid is said to have undergone a 'fragile-to-strong' (FS) transition. The term 'fragility' reflects the nomenclature of ordinary GF liquids for which liquids having a strong variation in their effective activation energy (slope in an Arrhenius plot) with $T$ are termed 'fragile', while those having a relatively weak $T$ dependence are termed 'strong'. Paradoxically, these GF liquids exhibit *both* of these behaviors in their respective high and low $T$ regimes of glass-formation. While such materials at low $T$ exhibit the rheological characteristics of a solid, justifying the term 'glass', these materials clearly have some features that set them apart from 'ordinary' GF liquids. Another singular feature of materials undergoing this type of FS glass-formation is that there is a well-defined peak in the specific heat $C_p$ and extrema in other thermodynamic response functions (thermal expansion coefficient, isothermal compressibility) that signals some kind of thermodynamic transition phenomenon (not a phase transition because the peak is too rounded) and no observable transition or only a weak specific heat $C_p$ feature arising from going out of equilibrium as in ordinary GF materials. In contrast, there is no observable or only a weak [144,145] thermodynamic signature of a 'liquid-liquid' transition in fluids exhibiting "ordinary" glass-formation, but, $C_p$ shows a large drop as the material goes out of equilibrium, a kinetic phenomenon that is normally taken as the *definition* of $T_g$.

We note that there has long been a discussion and controversy relating to a putative 'liquid-liquid' transition temperature $T_{LL}$ in polymer melts, and other 'ordinary' GF liquids, whose value is typically reported to be in the range $T_{LL} \approx$ (1.2 to 1.3) $T_g$. In particular, the weak theoretical rationale and subtle nature of the thermodynamic signature demonstrating the reality of such a transition has made this quantity controversial in the academic material science community, until recently, but the practical importance of this $T$ is broadly recognized in the field process engineering because this $T$ often signals gross changes in fluid flow and diffusion processes that are highly relevant for material processing. [146-153] We will see that GF fluids showing a FS transition exhibit a $C_p$ peak that follows a similar phenomenology in relation to its position relative to $T_g$, but the intensity of the thermodynamic feature, and the occurrence of anomalies in other thermodynamic response functions, indicate that this characteristic $T$ corresponds to a proper thermodynamic transition in the material. We take this as implying that there are possible unifying features in GF fluids exhibiting ordinary and FS glass-formation. Our investigation of dynamic heterogeneity in a metallic GF liquid exhibiting FS glass-formation further supports this essential unity in the physical nature of these materials, but it is not clear that any theoretical framework, or even correlative scheme, can encompass both classes of GF liquids. The theoretical challenge of developing a theory of the dynamics of GF liquids that encompasses both classes of materials then provides a good opportunity for improving our theoretical understanding of GF liquids broadly.

It is now clear after the discovery that water, silica and many metallic GF materials [12,94-96,154] follow the FS pattern of glass-formation that this is actually a rather common type of glass-formation. This brings us to the question of what essential physical features are shared by systems undergoing this type of glass-formation and which are distinct. A recent study suggests that the softness, a structure parameter identified by machine learning, provides a good description of FS transition in silica. [155] In particular, we may wonder whether the dynamic heterogeneities in these GF liquids take a similar form in these apparently distinct classes of GF liquids and whether models that have been applied to the description of diffusion and relaxation in ordinary GF materials also phenomenologically describe these other materials. Evidently, the usual phenomenological equations of relaxation and diffusion in ordinary GF liquids, such as Vogel-Fulcher-Tammann (VFT) equation, etc., and many others of more recent development [156-158], cannot describe the $T$ dependence of diffusion and relaxation in materials exhibiting FS glass-formation. Moreover, it is not clear whether the recently developed string model of glass-formation[7,16] can account for this pattern of relaxation so a comparison of simulations to this model would provide an important test of this model. We further point out that this general pattern in non-Arrhenius diffusion in fluids undergoing a fragile to strong transition is also observed in simulations and measurements on superionic *crystalline* materials. A simulation study of the O ion diffusion coefficient $D_O$ in superionic $UO_2$, [79] which is remarkably high for a crystalline material, explaining the term 'superionic', indicted that the string model of glass-formation could quantitively describe $D_O$ in this material at all $T$ simulated so the applicability of the string model to describe GF liquids undergoing a FS transition cannot be ruled out a priori.

**Appendix B: How do we explain the origin of the fragile-strong transition?**

Although it is nice to know that the relation between collective motion and changing activation energy still applies to our metallic glass material undergoing a FS transition, we still have not answered why glass-formation in this material, and many metallic and others materials [159], follows such a different pattern of glass-formation thermodynamics and dynamics. Another feature of materials undergoing this type of FS glass-formation is that there is a well-defined peak in the specific heat $C_p$ and extrema in other thermodynamic response functions (thermal expansion coefficient, isothermal compressibility), signaling some kind of thermodynamic transition phenomenon. On the other hand, there is no observable or only a weak [144,145] thermodynamic signature of such a 'liquid-liquid' transition in fluids exhibiting 'ordinary' glass-formation. The hypothesis that systems exhibiting such thermodynamic transition features are exhibiting a kind of cooperative thermodynamic self-assembly transition [1,16,160] offers an attractive possibility for interpreting such systems. It is characteristic of materials exhibiting FS transitions to exhibit 'anomalies' in thermodynamic response functions [40,57,161-163] such as the specific heat, thermal expansion coefficient of the fluid, isothermal compressibility that accompany sharp changes in the viscosity and viscoelasticity of solution exhibiting supramolecular assembly (See appendix A). These thermodynamic anomalies are well known in the case of water [96,164], where they have often been attributed to a counter-intuitive liquid-liquid phase separation of water from itself, [165] a phenomenon arising from water's capacity to exhibit a multiplicity of distinct local packing states having distinct thermodynamic and dynamic signatures- mobile high density water and low density immobile water. [96] It is then a matter of interest to consider whether these thermodynamic changes in response functions also arise in our metallic glass material to see if we can find any general thermodynamic clues about the nature of glass-formation in this class of materials. Accordingly, we estimated the specific heat of our metallic glass material and the four-point density correlation $\chi_4$ and find that this system indeed exhibits well defined anomalies of a thermodynamic nature.

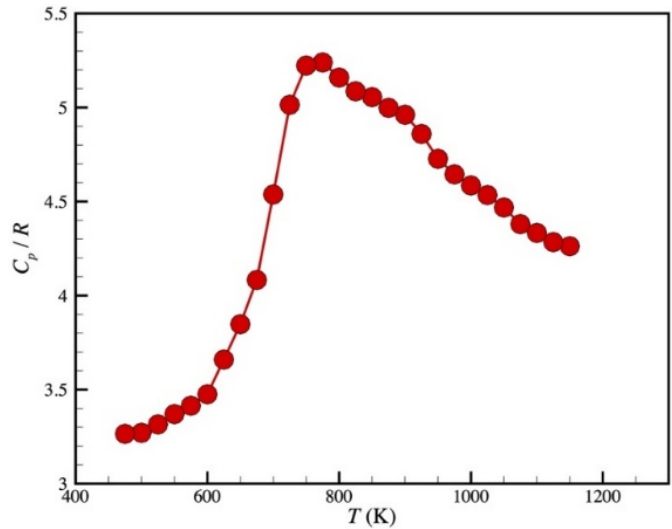

**Figure 20.** Observation of a lambda transition in the specific heat. Specific heat $C_p$ at constant pressure for our $Al_{90}$ $Sm_{10}$ metallic GF liquid as a function of $T$. We see that $C_p$ normalized by the gas constant $R$ exhibits a peak around 750 K, a temperature not far from the lambda transition temperature $T_\lambda$ identified in the differential activation energy plot shown in Fig. 1(b). Superionic crystalline materials also exhibit a rounded specific heat anomaly consistent with a rounded thermodynamic transition. [166] Note that this peak is not the result of falling out of equilibrium, but rather represents a true thermodynamic transition of some kind. See Refs. [95], [96], and [164] for discussion of the lambda transition in water.

We first consider the specific heat at constant pressure $C_p$ in Fig. 20 where we see that $C_p$ exhibits a rounded peak, consistent with a mildly cooperative self-assembly transition.[161] It is characteristic of such self-assembly transitions that they are *'rounded'* so that are no singularities exist in the free energy that give rise to divergence in any response functions, as found in second order phase transitions.[167] Next, we consider the 4-point density autocorrelation function which should provide information about density fluctuations accompanying self-assembly. To define this quantity, first, we define time-dependent self-overlap function $Q_s(t)$,[168]

$$Q_s(t) = \sum_{j=1}^{N} \omega\left(|r_j(t) - r_j(0)|\right) \quad (7)$$

when $|r_j(t) - r_j(0)| < 0.3\sigma$, $\omega = 1$, and when $|r_j(t) - r_j(0)| \geq 0.3\sigma$, $\omega = 0$, where $\sigma$ is the average atomic spacing of a particular system and 0.3 is chosen as a typical amplitude of caged particles. The detailed description on the choice of cut-off has been provided elsewhere. [11] The mean-squared variance of $Q_s(t)$ then defines the (self-part) of $\chi_4$,

$$\chi_4 = \frac{V}{N^2}[\langle Q_s(t)^2 \rangle - \langle Q_s(t) \rangle^2]. \quad (8)$$

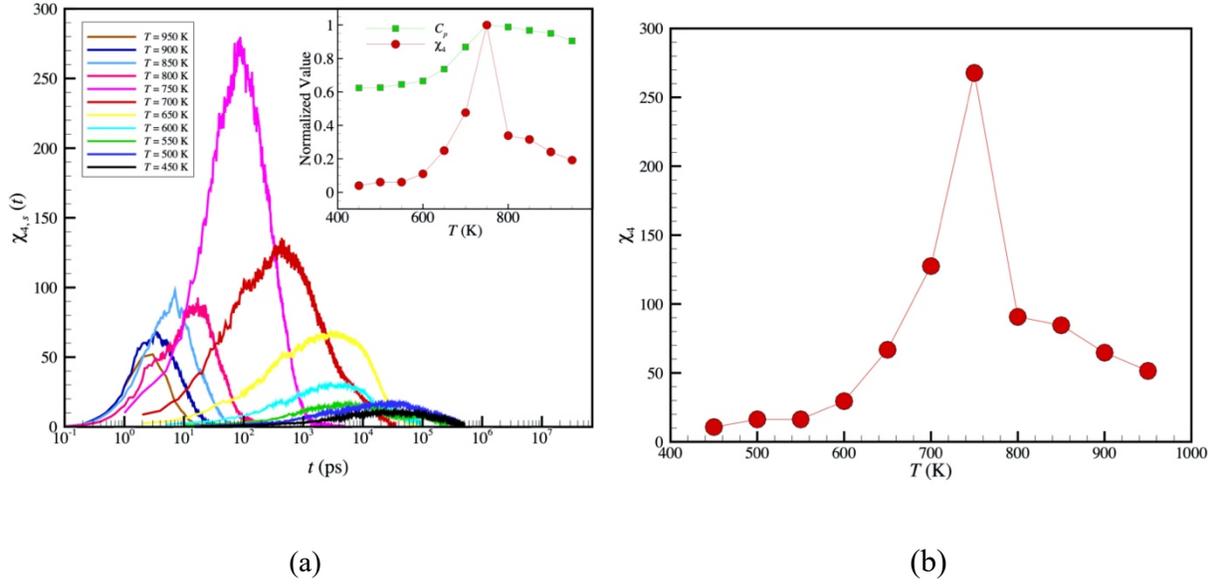

(a)            (b)

**Figure 21.** Four-point density correlation function as a function of $T$. (a) Four-point density correlation function $\chi_4(t)$ (left figure) and its peak value (right figure) for our $Al_{90} Sm_{10}$ metallic GF liquid as a function of $T$. The peak in $\chi_4(t)$ occurs near the lambda transition temperature $T_\lambda$ at which $C_p$ exhibits a peak (see discussion of this $T$ and other characteristic temperatures in SI) The inset of the left figure directly compares the $T$ variation of $C_p$ and the peak in $\chi_4(t)$. (b) Peak in $\chi_4$ as a function of $T$. For a discussion of this thermodynamic feature in the case of water; see Refs. [95], [96], and [164].

We see in the Fig. 21(a) that $\chi_4(t)$ exhibits a qualitative variation that is similar to the number of particles of high and low mobility and, following the usual convention, we define the time at which $\chi_4(t)$ peaks as being a characteristic time of the material. As in the case of $C_p$, we see that this response function peaks around 750 K, which is again in the vicinity of the lambda transition temperature $T_\lambda$ defined above.

The temperature derivative of the density $d\rho/dT$, or more specifically, $-(d\rho/dT)/\rho$ defines the thermal expansion coefficient and we see in Fig. 22(a) that this quantity also has a peak near $T_\lambda$. The density $\rho$ as a function of $T$, shown in the insert to Fig. 22(a), correspondingly has a kink near this characteristic temperature, a thermodynamic feature characteristic of fluids exhibiting equilibrium polymerization. The weak nature of these singular features is characteristic of highly rounded thermodynamic transitions, [40,161-163] as one finds in the Ising model with a strong applied magnetic field. For completeness, we also calculate the isothermal compressibility,

$$\kappa = (<V^2> - <V>^2)/(k_B T <V>), \qquad (9)$$

as function of $T$ where $V$ is the volume of the system and $<...>$ denotes NPT ensemble average. This response function likewise has a peak near $T_\lambda$, but the strength of this feature is rather weak, and, moreover, when we combine $\kappa$ and $\rho$ to estimate structure factor in the long wavelength limit, $S(0) = \rho\,\kappa\,k_B\,T$, we find (results not shown) only a smoothly varying function of $T$ having an *inflection point* near $T_\lambda$. The osmotic compressibility of solutions exhibiting equilibrium polymerization correspondingly exhibits a sigmoidal variation with $T$. [161] We finally remark that all these thermodynamic response functions exhibit singular features that are much stronger in the case of supercooled water, although observations of these transitions in real bulk water are made difficult by the strong tendency of water to crystallize near the $T$ where measurements of confined water and simulations indicate a peak in the specific heat and the other response functions discussed above. [95,96] We next provide a tentative physical explanation of the observed growth of dynamic heterogeneity in GF liquids and associated thermodynamic properties just described based on established models of the thermodynamics of materials composed of associating particles.

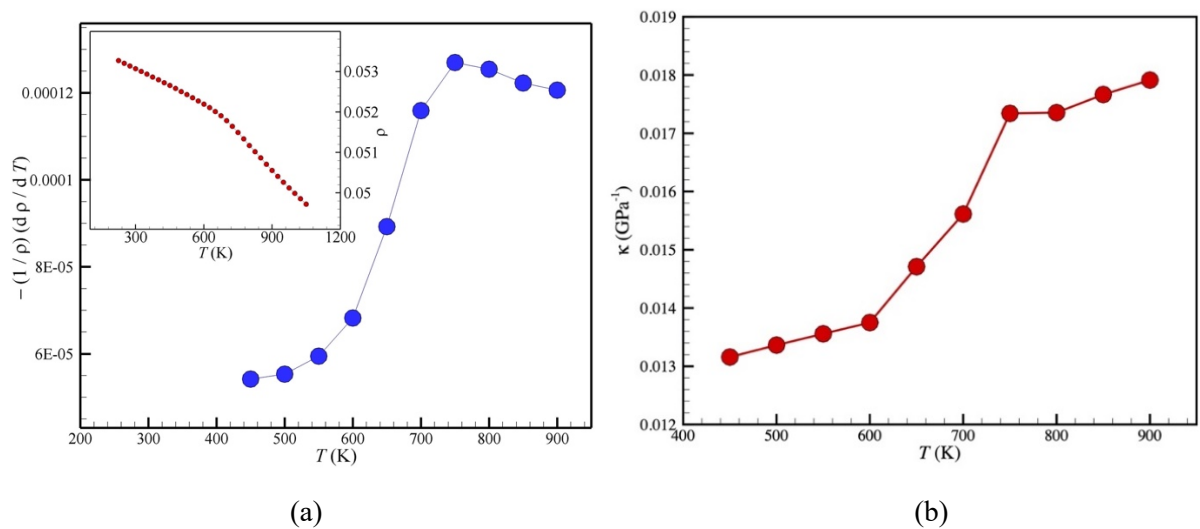

(a)           (b)

**Figure 22.** Observation of thermodynamic signatures in the temperature dependence of the density and compressibility. (a) Temperature derivative of the density $\rho$ or thermal expansion coefficient for our $Al_{90}$ $Sm_{10}$ metallic GF liquid as a function of $T$. (b) Isothermal compressibility for our $Al_{90}$ $Sm_{10}$ metallic GF liquid materials as a function of $T$. Notice the similarity of the kink in the density near $T_\lambda$ to a feature observed by Fox and Flory[1] in the density of polystyrene melts at a $T$ later identified as the 'liquid-liquid' transition temperature.[9]

Douglas and coworkers [1,16,97,160] have previously argued that glass-formation can viewed generally as a particular kind of supramolecular assembly transition in the formation of the form of dynamic polymers, i.e., strings. This should not be surprising since many second order phase transitions [equilibrium polymerization, Ising model, XY model, Heisenberg model, spherical model] can be formulated in terms polymerization transitions with different weights on the self-interactions of clusters describing correlations in these systems [169-171] and Douglas and Ishinabe [172] used this formal theoretical framework to estimate the critical temperatures of this important class of spin models based on geometrical information about self-avoiding paths. Moreover, the problems of thermal initiation of equilibrium polymers or activated polymerization as in the polymerization of actin and sulfur upon change of temperature can be formulated in terms of an Ising model with an applied magnetic field to model the initiator or activation processes regulating the polymerization process. [173,174] Quite apart from this abstract perspective, we actually see dynamic structures in our simulations of ordinary GF liquids (both metallic glasses and polymeric) consistent with activated equilibrium polymerization. This point of view has been discussed in previous papers aimed at understanding ordinary glass-formation [1,16,97,160] and here we consider if this type of treatment with the present metallic GF material.

We note at the outset that the observation of sharp features in thermodynamic response functions such as the specific heat, density and osmotic compressibility are characteristic of equilibrium polymerization transitions that are highly cooperative, [161] as in the well-known case of the equilibrium polymerization of sulfur, where a lambda transition, again defined by a peak in $C_p$. [57] Exactly solvable models show that sharpness of the rounded polymerization transition [40,162] when varying temperature and magnitude of $C_p$ can be tuned over a wide range by varying the initiator concentration, rate of activation or other physical processes that initiate the polymerization process. [163] Douglas and coworkers [161] suggest that the degree of cooperativity of this class of transitions is directly related to the extent that the assembly process resembles a phase transition. It thus seems very likely that this perspective might encompass glass-formation of the ordinary type glass-formation for which the transition is highly rounded and the equilibrium specific heat feature is not apparent to highly cooperative types of glass-formation for which strong thermodynamic anomalies are apparent at characteristic temperatures to define the transition. In SI, we provide detailed description and discussion on how we determine the

65characteristic temperatures of glass-formation ($T_o$, $T_g$, $T_c$, $T_A$), the lambda transition temperature, $T_\lambda$, the fragile-strong transition temperature, $T_{FS}$, and the bifurcation temperature, $T_{\alpha\beta}$. It is also characteristic of this type off rounded thermodynamic transition that there is definite onset temperature where the association begins appreciably and where the transition saturates after the structures have fully formed after sufficient cooling. It is also notable that the inverse scaling between the average size of the exponentially distributed polymer structures can be derived to scale inversely to the configurational entropy of the fluid[162], as AG assumed heuristically.

One of the most basic properties to consider in the equilibrium polymerization is the fraction of the particles that are in an assembled polymeric state of all particles that are free to possibly associate. This is the order parameter of the self-assembly process and we may determine this order parameter for the string assembly process discussed above. In Figure 23, we show the fraction of mobile particles that are organized into the string state, $\Phi$. We observe that the onset temperature $T_A$ is close to the $T$ at which $\Phi$ starts to increase upon cooling, the peak of $d\Phi/dT$ defining the locus of the polymerization transition [161] is close to $T_\lambda$, where the specific heat transition occurs in both the polymerization model and our metallic glass simulations and the "saturation temperature" [160] where the self-assembly transition is over around 450 K. In the polymerization model, the saturation temperature corresponds to a $T$ at which the configurational entropy of the fluid associated with the self-assembly process starts to flatten off as the assembled structures become completely organized and we are tempted to call this a thermodynamic 'glass transition temperature' $T_g$ since there is phenomenological evidence that the fluid configurational entropy in GF liquids becomes slowly varying below the calorimetric $T_g$ of measurements under cooling conditions where a very slow cooling rate is employed. The significance of this $T_g$ in the self-assembly model is that the self-assembly transition is simply 'over' at this point. All these characteristic temperatures of the FS glass-formation process are reflected directly in the differential change of the activation energy $E_{diff}$ in Fig. 1(b) and the $T$ dependence of the 'polymerization order parameter' $\Phi$ resembles the $T$ dependence of the true activation energy in Fig. 1(a). We note that Tanaka and coworkers [96] have introduced a two-state model of water, which is similar in spirit to the polymerization model, except this model does not consider the spatial correlation between 'ordered' species. In this model, the fraction of particles converted to a thermodynamically favored state of different mobility which plays a key role in their modeling of the activation free energy. Douglas et al. [161] have shown this type of two state serves as an approximation to more complex supramolecular association models (the state of the mobile particles be considered in abstract sense to be either in a polymerized or unpolymerized state if you have no concern about the cluster connectivity) for the purposes of estimating $\Phi$. Cooperativity in this simple type of kinetic model can be tuned by formally changing the order of this reaction to match the cooperativity exhibited by polymerization process involving a chain of reactions involved in forming the polymer. [161] Our main point here is that modeling of Tanaka and coworkers of the dynamics and thermodynamics of water [96] can be considered to be very much in the same spirit as the equilibrium polymerization model and this approach likewise leads to anomalies in the thermodynamic response functions, as seen in the simulations of both water and our own simulations. We expect the more complex polymerization model to become more useful when subjecting fluids to flow or amorphous solids to deformation where the deformation will interact and alter the structure of the self-assembled clusters, thereby influencing the dynamics of the GF material. [2]

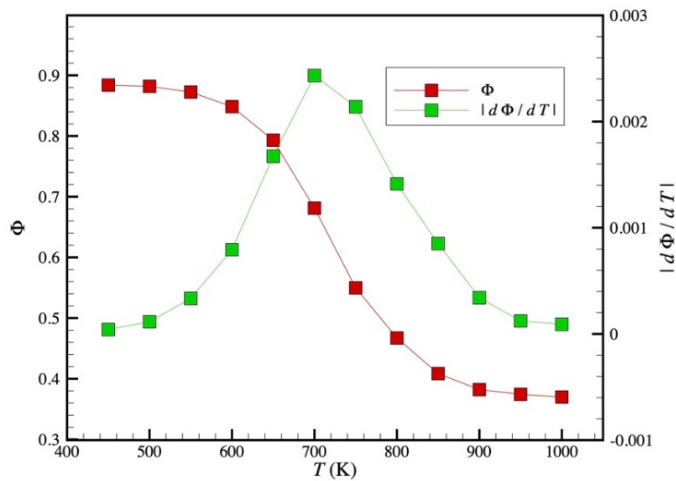

**Figure 23.** String order parameter as a function of *T*. Order parameter Φ for the polymerization of clusters of chains exhibiting string-like collective motion for our $Al_{90}Sm_{10}$ metallic GF liquid as a function of *T*. We also plot the derivative Φ, where we notice the similarity of these quantities to the variation of the activation energy in Fig. 1(a) and the derivative dΦ / dT to the differential activation energy in Fig. 1(b).

We mention that the activation free energy was found previously to vary sigmoidally with *T*, as we have found in the present metallic glass material, when fit our simulation data for the segmental relaxation dynamics for a coarse-grained polymer melts quantitatively to the string model with an initiated equilibrium polymerization.[1,16] Notably, the polymer melt exhibited the conventional pattern of glass-formation discussed above [16], and thus did not exhibit a FS transition. The difference between this conventional polymeric glass-forming liquid and the present metallic glass-forming liquid is the *broadness of the transition*. When the glass transition is much broader, the relaxation time grows so much during the course of cooling that the system just can't complete the thermodynamic crossover between the high and low *T* Arrhenius regimes because of a non-equilibrium structural arrest. The relatively rapid rate at which systems exhibiting a FS type glass-formation "complete" the thermodynamic transition allows the full sigmoidal variation of the activation energy to be observed. We think that variations in the degree of cooperativity of the glass transition then accounts for the different classes of GF liquids and that all GF liquids should be describable in terms of a unified self-assembly based description in which the source of this variable cooperativity is addressed. This remains to be seen, but this point of view of glass-formation liquids looks promising.

**Appendix C: Equilibrium Polymerization, Liquid-Liquid Transition and Glass-Formation**

We mention that the idea that equilibrium polymerization underlies glass-formation can be traced back to the earliest modeling of GF liquids. In 1934, Hägg [175] responded to Zachariasen's introduction of random network model of glasses[176] by suggesting that both metallic and inorganic glasses involved the formation of one-dimensional chains (or two-dimensional sheet polymers or highly perforated sheets, i.e., branched polymers[110]) often composed tetrahedra in ion complexes or arising from bonding habits of low coordination in 'metalloid' elements such as the chalcogenides, first forming well into the liquid state, but growing progressively upon cooling, thereby frustrating crystallization because of difficulty of molecular species localized in clusters to diffuse and by the "muddle of other chains" that further inhibit cluster movement, i.e., entanglement. Apart from the prescience of this model of glass-formation it was remarkable that the concept of polymers was hardly accepted at this time. After a heated exchange between Zachariasen and Hägg, [177,178] Regrettably, Hägg's model sank into oblivion until recently where measurement have largely confirmed his original conception of a specific form and origin of a specific form of dynamic heterogeneity found in many GF liquids.

The polymerization model of glass-formation, in which the 'ordering' process in many GF liquids is viewed as a form of equilibrium polymerization [97,160], also has strong implications for the miscibility of metal alloys and other multicomponent glass-forming liquids since polymerization inherently alters the miscibility of mixtures and can lead to a multiplicity of critical points arising from competition of the polymerization thermodynamics and phase separation[163,179,180] [See Figs 6a and 6b and ref. [163]] and the change in miscibility due this physical cause is particularly significant when one of the components self-assembles upon heating, as in the case of S. This type of additive is predicted [181] to give rise to closed loop phase behavior, as observed in S-Te mixtures. [182-185] This type of phase behavior is well known and Tsuchiya [179] recognized the relevance of this type of 'structural change' in liquid mixtures on the thermodynamic stability of Te alloys. This phenomenon has been well-studied in related problem S solutions [57] (See Fig. 1a of Ref. [57]) and in the phase diagram of $He^3$-$He^4$ mixtures [186,187], systems which are also mentioned by Tsuchiya. [179] Note that the liquid-liquid transition between the normal and superfluid liquid states occurs as a line of second order phase transitions emanating from the critical point in the He mixtures can be viewed as a polymerization line based on Feynman's equilibrium polymerization model of the superfluid transition where the polymers were assumed to move collectively in permutational motion having the form of polydisperse polymer rings. [188,189] We will see this type of permutational motion in connection with the classical molecular dynamics of our Al-Sm metallic GF liquid. This type of phase behavior, involving a line of polymerization or self-assembly line of transitions that terminates near the critical point of the phase boundary for liquid-liquid phase separation, is ubiquitous in complex fluids in which molecular assembly occurs, micelle formation [190,191], living polymers [57], thermally reversible gelation of polymers in solution [192,193], etc. Liquid-liquid thermodynamic transitions, accompanied by significant changes in fluid structure and viscoelastic properties in the two liquid states, are thus very common in complex fluids, even if this phenomenon is not always overtly recognized. In our companion paper focusing on the fast dynamics of our Al-Sm metallic glass [4], we observe clear evidence of density fluctuations at a nanoscale, providing some evidence of a liquid-liquid phase separation of the metallic glass into two distinct density states. Our observations accord with the findings of the combined experimental and computational study of $Cu_{64}Zr_{36}$ metallic glasses by Feng et al. [194]

From a practical standpoint, we note that FS glass-formation plays an essential role in the practical performance of phase-change memory materials (compounded from Te and other chalcogenides and group-IV and group-V elements [195] to which the chalcogenides strongly bind) because it allows for rapid switching between crystalline and amorphous states in connection with the speed of data recording, in addition to 'non-volatility' of stored information[196,197] because of the ultrastable nature of this class of materials in their glass state. [46,198] The accompanying large changes in conductivity and other properties accompanying this structural transition also explain the strong attraction of these materials for applications. In the metallic glass context, this structural transition is accompanied by a significant increase ($\approx 20$ %) of the hardness of the material. [46,199] Extensive evidence for liquid-liquid phase separation in chalcogenide-based phase-change materials. [200] Particularly informative measurements of the singular variation of $C_p$ and other response functions for Te and alloys with other chalcogenide atomic species have been described by Tsuchiya [64,65] Recent first principle quantum mechanical calculations for cooled water [201] have also indicated the formation dynamic polymeric water clusters that are rather reminiscent of those just described in metallic systems so it is not surprising that all these systems exhibit FS type of glass-formation accompanying this type of structural organization process. Kanno et al. [202] have drawn a strong analogy between the structural transformation and the associated lambda transition temperature in chalcogenides and the thermodynamic anomalies of water.

## Acknowledgements

H.Z. and X.Y.W. gratefully acknowledge the support of the Natural Sciences and Engineering Research Council of Canada under the Discovery Grant Program (RGPIN-2017-03814) and Accelerator Supplements (RGPAS-2017- 507975).

## Data Availability Statements

The data that supports the findings of this study are available within the article and its supplementary material.

## Supplementary Material

See supplementary material for Simulation Sample Preparation, Cooling Rate, and Equilibration, determination of diffusion coefficients, determination of characteristic temperatures of glass-formation and the lambda transition temperature, quantifying immobile particle cluster size, and interpretation of $\tau_\alpha$ as a structural relaxation time.

# Supplementary Information:

# Dynamic Heterogeneity, Cooperative Motion, and Johari-Goldstein $\beta$-Relaxation in a Metallic Glass-Forming Material Exhibiting a Fragile to Strong Transition

Hao Zhang[1†], Xinyi Wang[1], Hai-Bin Yu[2], Jack F. Douglas[3†]

[1] Department of Chemical and Materials Engineering, University of Alberta,

Edmonton, Alberta, T6G 1H9, Canada,

[2] Wuhan National High Magnetic Field Center, Huazhong University of Science and Technology, Wuhan, Hubei, 430074, China

[3] Material Measurement Laboratory, Material Science and Engineering Division, National Institute of Standards and Technology, Gaithersburg, Maryland, 20899, USA,

[†]Corresponding authors: hao.zhang@ualberta.ca; jack.douglas@nist.gov

## 1. Simulation Sample Preparation, Cooling Rate, and Equilibration

In a previous study, centred Sm atoms with a '3661' short range order (SRO) structural motif were identified as the most abundant SRO unit in $Al_{90}Sm_{10}$ metallic glass system based on ab initio molecular dynamics (MD) simulation, following a procedure utilized before in our previous Voronoi tessellation analysis of a $Cu_{64}Zr_{36}$ metallic glass. [1] More recently, MD simulations were employed to investigate the dependence of this SRO motif on the cooling rates, where the liquid $Al_{90}Sm_{10}$ specimen was continuously cooled from 2000 K to 300 K with five different cooling rates, i.e., $10^{13}$, $10^{12}$, $10^{11}$, $10^{10}$, and $10^{9}$ K/s. [2] It was found the population of '3661' Sm centred motif increased sharply upon $T_g$ and then gradually approached to a constant value at low $T$ region, and the total population of the '3661' motif was sensitive to the cooling rate, i.e., the lower the cooling rate, the larger the population of this motif. The behavior of '3661' SRO clusters, where the cluster population increases sharply approaching upon glass transition temperature $T_g$, is rather similar to the behavior of icosahedral clusters associated with SRO in Cu-Zr metallic glasses. However, such SRO motifs do not tend to form interpenetrating network even at $T$ well below $T_g$.

In the current study, in order to obtain a well 'equilibrated' $Al_{90}Sm_{10}$ materials at low temperatures, we applied an even lower cooling rate (one order lower than the slowest cooling rate used in the above study[2]), i.e., a cooling rate of $10^8$ K/s from 2000 K to 200 K, which takes a total of 18,000 ns simulation time. Although an extremely low cooling rate was employed in the current study, we expect that the system still would not reach complete equilibrium at low temperatures, especially when the Johari-Goldstein relaxation process is considered at relatively low $T$. Note that although our simulations are often on the order of a µs timescale, this time is still rather short in comparison with the structural relaxation time $\tau_\alpha$ near $T_g$, which is on the order of a minute. This is an inherent limitation of molecular dynamics simulations which has largely precluded investigation of the Johari-Goldstein relaxation process. Nevertheless, this is the best we can achieve under the current available computing power and in Figure S1, we exhibit that the temporal evolution of the average value of potential energy and number density of the system do not change visibly with time at $T$ = 450 K, 500 K, and 550 K, which suggest a near equilibrium or at least highly stable metastable condition has been achieved in the time window of our observations.

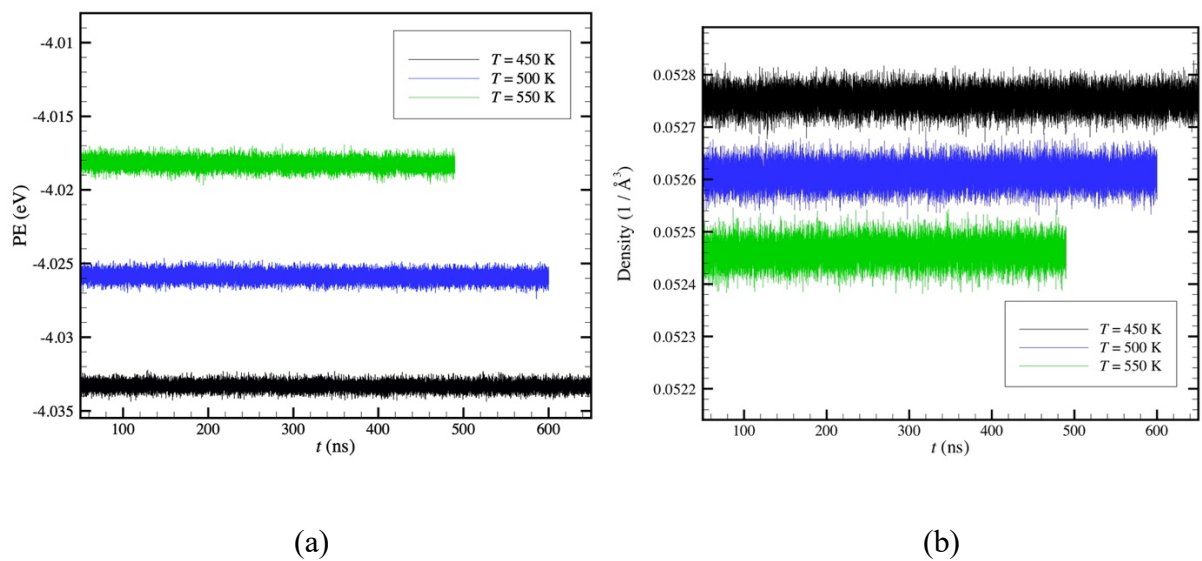

(a)          (b)

**Figure S1.** Quantification of potential energy and density fluctuations as a function of time. (a) The average potential energy per atom as a function of simulation time at $T$ = 450 K, 500 K, and 550 K. (b) The number density of the system as a function of simulation time at $T$ = 450 K, 500 K, and 550 K. Both quantities do not show drift with time in the simulation window we examined.

## 2. Determination of Diffusion Coefficients, *D*

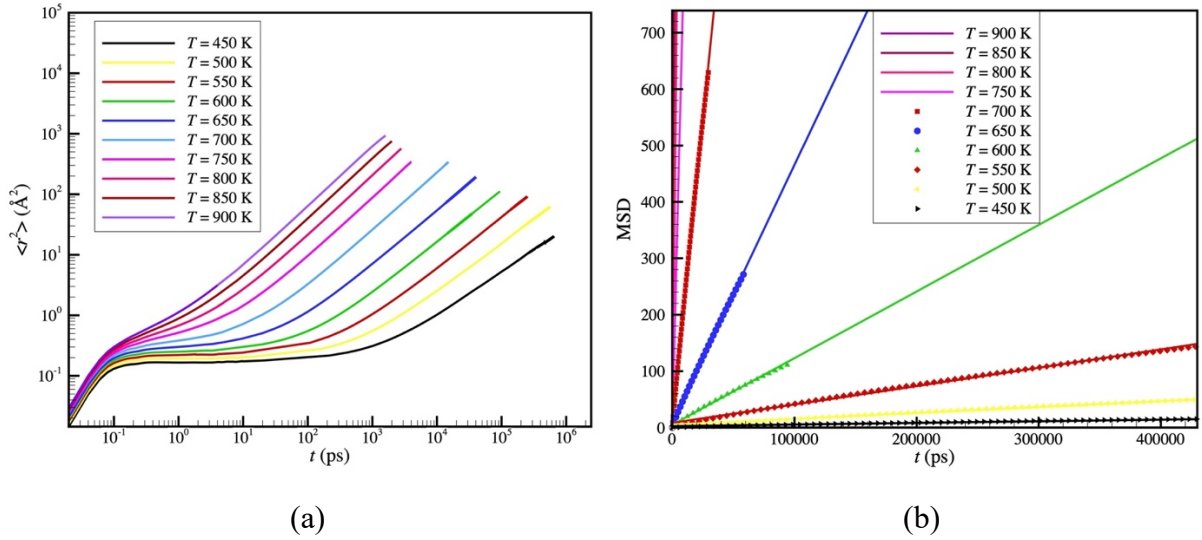

(a)     (b)

**Figure S2.** Average mean square atomic displacement in time and the average diffusion coefficient, *D*. (a) Average mean square displacement of atoms in an $Al_{90} Sm_{10}$ metallic glass over a range of *T* in a log-log scale. *D* is obtained from the slope of these curves at long times in (b).

A standard computational procedure is used to determine the diffusion coefficient, i.e, $D = \frac{1}{6} \lim_{t \to \infty} \frac{d<\Delta r^2>}{dt}$, $<\Delta r^2(t)> = \frac{1}{N}\sum_i (\mathbf{r}_i(t) - \mathbf{r}_i(0))^2$, in the limit of long times, where $\mathbf{r}_i(0)$ and $\mathbf{r}_i(t)$ are particle's initial and final atom positions after time *t*, respectively, and *N* is the total number of atoms. In Fig. S2 (a) we show the mean square displacement $<r^2(t)>$ of all the atoms in our system in short time regime in which the 'caging' of all particle motion in an intermediate time between the short time regime where $<r^2(t)>$ grows linearly in time. Figure S2 (b) shows $<r^2(t)>$ over a much larger timescale where the caging behavior appears just as a short time transient for the *T* range shown. Section 3 below discusses the estimation of the Debye-Waller Factor $<u^2>$, which is just $<r^2(t)>$ evaluated at a precisely defined 'caging time'; See Fig. S4.

*D* is obtained in a standard fashion from the slope of these curves at long times. The average diffusion coefficient *D* is defined as the *composition weighted* (atomic fraction) average of the Al and Sm components and our estimates of *D* are shown in Fig. 1 a) of the main text. See Douglas et al. [3] for a discussion of *D* and the atomic component values of *D* values for Cu-Zr metallic glasses having a range of compositions.

For completeness, the diffusion coefficients of each atomic species ($D_{Al}$, $D_{Sm}$) are shown in Figure S3 as a function of $1/T$ where the data is shown on a semi-log scale because of the large range of the data. We observe from Fig. S3 that the average *D* for the metallic glass practically coincides with $D_{Al}$ and similarly $\tau_\alpha$ for the average structural time of the entire material is roughly equal the structural relaxation time of the Al component $\tau_\alpha(Al)$, as apparent in Fig. 3 of the main text. The relation between the component and the average $\tau_\alpha$ is discussed and illustrated at length by Douglas et al. [3] for the Cu-Zr metallic glass so our discussion here is brief. It is evident from Fig. S3 that the diffusion coefficient of the smaller atomic species (Al) can be orders of magnitude larger than the larger species (Sm). Moreover, Fig. S3 and Fig. 3 of the main text indicate that the average diffusion coefficient and structural relaxation time are both dominated by the fast-moving smaller atomic molecular species. This greatly accelerated dynamics arising from smaller atomic species has been observed experimentally in radio isotope measurements on metallic glass materials.[4]

The direct relation between the average *D* and the JG *β*-relaxation time that we have noted in the main body of our paper further implies from the approximation $D \approx D_{Al}$ that the JG *β*-relaxation process should be correspondingly dominated by the rate of diffusion of the smaller atomic species. Measurement studies on model metallic glasses have correspondingly indicated a strong correlation between the diffusivity of the smaller atomic species and the JG *β*-relaxation time. [5] We note that this is a matter of some practical importance in view of the significant importance of the JG $\beta$-relaxation process in relation to the brittleness and other basic mechanical properties of metallic glass materials. [6] The extreme mobility anisotropy in the present material can be naturally understood physically from the rather large difference in the atomic size of Al and Sm. Al as an atomic radius of 1.25 Å, while Sm has an atomic radius 1.85 Å defined in terms of a standard empirical measure of atomic radii [7] so that the Sm atoms are roughly 50% larger than Al atoms, which is a size anisotropy that we may fairly characterize as being 'unusually large' in magnitude. The strong dependence of $D$ on atomic species is well-known in metallic GF materials [8] and an extensive review on this topic has been given by Faupel et al. [9] (See Sect. V), which also discusses the various estimates of atomic 'size'.

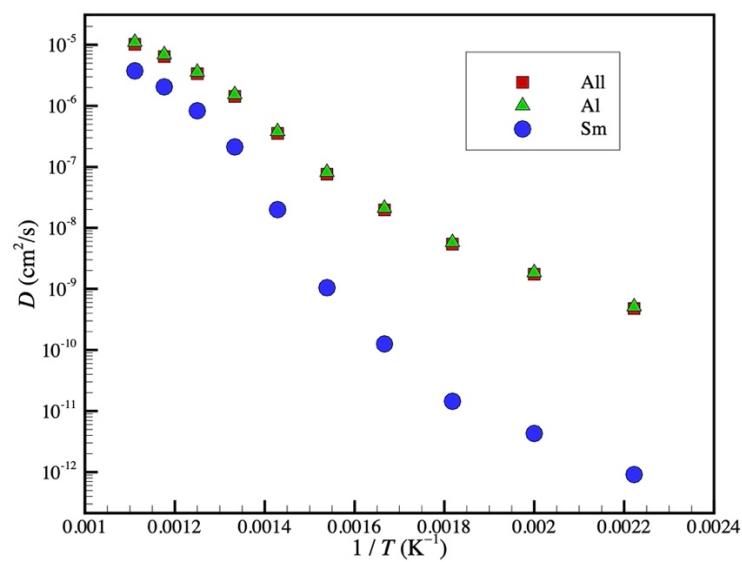

**Figure S3.** Diffusion coefficients of Al and Sm atoms for an $Al_{90}Sm_{10}$ metallic glass over a range of $T$ plotted on a semi-log scale. The diffusion coefficient $D$ of each atomic species is obtained from the slope of these curves at long times based on plots similar to the data shown in Fig. S2.

This discussion of chemically-specific aspects of the dynamics of the metallic glass system that we study brings us to finally consider why the JG $\beta$-relaxation process is so well separated from the $\alpha$-relaxation process in this material, a situation that normally allows to study the JG $\beta$-relaxation process without the strong overlap with the $\alpha$-relaxation process normally observed in metallic glasses. This is a difficult question to answer with absolute certainty, but we suggest that the exceptionally large dynamic asymmetry that accompanies the large atomic size asymmetry, noted above, is responsible for this large timescale separation. We previously found in simulations of Cu-Zr metallic glass materials that the degree of 'decoupling' between $D$ and the structural relaxation time $\tau_\alpha$ correlated strongly with the fragility of glass-formation and relative mobility of the atomic species, as quantified by the Debye-Waller factor. [3] A large decoupling means a large separation between the structural relaxation time and the characteristic hopping time associated with particle diffusion, i.e. relatively accelerated diffusion. The predominance of the average rate of diffusion in the material by the smaller particles having higher mobility means that smaller particles should give rise to a greater separation between the $\alpha$-relaxation process and the JG $\beta$-relaxation process, provided JG $\beta$-relaxation process is generally controlled by the rate of atomic diffusion, as seen in our simulated Al-Sm material. This simple interpretation of the timescale separation is probably not completely general, however, because there is abundant evidence that the strength of the JG $\beta$-relaxation process in both metallic and non-metallic materials is also dependent on the strength of the interparticle interaction, molecular stare of dispersion, nanoscale confinement, etc.[10-18] . Thus, we may not expect any simple geometrical interpretation of the strength of the JG $\beta$-relaxation process. Despite the complexity of trying to understand the JG $\beta$-relaxation process in GF materials broadly, it does appear that atomic or particle size generally is a relevant parameter for tuning the intensity of this relaxation process, along with the strength of molecular cohesion of particles with those in their surroundings. [10] Further work is needed to determine if dynamical asymmetry of the component species of GF systems can explain the separation in timescales between the JG $\beta$-relaxation process and $\alpha$-relaxation processes.

## 3. Determination of Characteristic Temperatures of Glass-Formation ($T_o$, $T_g$, $T_c$, $T_A$) and the Lambda Transition Temperature, $T_\lambda$

### A. Rough Estimates of Characteristic Temperatures from the Debye-Waller Factor $\langle u^2 \rangle$

The estimation of characteristic temperatures of glass-formation is more difficult for materials exhibiting a fragile-to-strong (FS) transition because one cannot rely on the standard phenomenology of glass-forming (GF) liquids such as the Vogel-Fulcher-Tammann (VFT) relation [19-21] or the conventional phenomenological definition that the α-relaxation time $\tau_\alpha$ at the glass transition temperature $T_g$ equals 100 s. As an example, we consider the situation of water, the most extensively studied fluid exhibiting a FS transition. [22,23] Estimates of $T_g$ for water based on the "100 s rule" or its shear viscosity equivalent in conjunction with the VFT relation, have indicated a $T_g$ estimate for water near (162 ± 1) K, [24,25] while other estimates based on specific heat $C_p$ measurements of amorphous ice have indicated a much value of $T_g$ near 136 K. [26] While the lower $T_g$ estimate seems to currently have a greater acceptance in the scientific literature, there is currently no general consensus on the value of $T_g$ for water, and one routinely finds both of these $T_g$ estimates as the 'true' value of $T_g$ of water. Below, we will find the same ambiguity plagues estimate of $T_g$ and $T_o$ in our metallic glass. We note that in recent years it has been realized that many GF liquids follow the FS pattern of glass-formation, [27] although this phenomenon seems to be especially prevalent in metallic GF liquids. [28]

In our past work, we found that we could find at least good rough estimates of the characteristic temperatures of glass-formation through a consideration of the $T$ dependence of the mean square displacement $\langle u^2 \rangle$ at a caging time scale, i.e., the Debye Waller Factor (DWF) defined as, $\langle u^2 \rangle = \langle \frac{1}{N}\sum_{n=1}^{N}\{(x_1 - x_0)^2 + (y_1 - y_0)^2 + (z_1 - z_0)^2\}\rangle$, where $(x_0, y_0, z_0)$ and $(x_1, y_1, z_1)$ are particle's initial and final positions after time $t = 1$ ps, respectively, [3,29,30] which is notably an equilibrium fluid property defined on a ps timescale. We first consider this method of roughly estimating the characteristic temperatures of glass-formation, which is admittedly untested previously for systems exhibiting a FS transition.

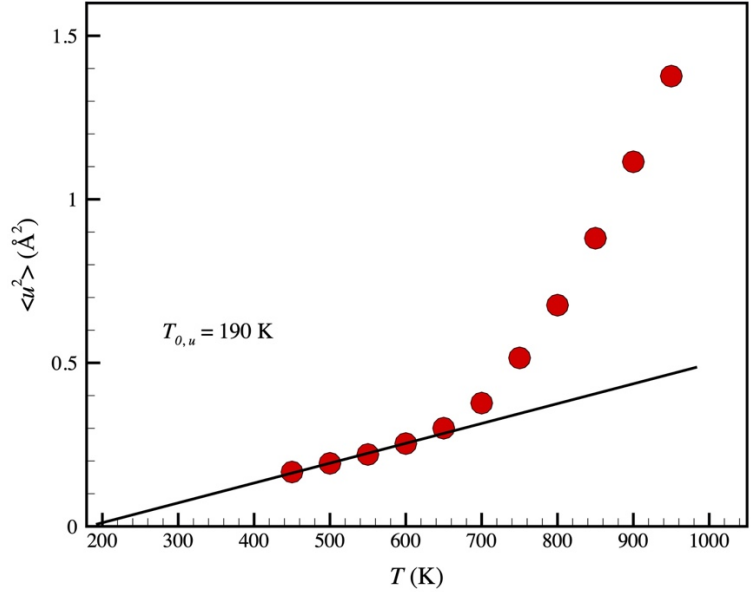

**Figure S4.** Temperature dependence of $\langle u^2 \rangle$ of our SM-Al metallic glass-forming liquid.

In Figure S4, we plot $\langle u^2 \rangle$ over the full $T$ range that we have investigated. We notice that usual, there is a low $T$ regime and that $\langle u^2 \rangle$ extrapolates to zero a finite $T$, defining a characteristic temperature, $T_{o,u} = 190$ K, indicating the 'termination' of the glass-formation process. It should be appreciated that this $T$ involves a long extrapolation and should not be interpreted as a $T$ at which $<u^2>$ literally vanishes given that our data is limited to a high $T$ regime. Within the Localization Model, [3] the characteristic temperature $T_{o,u}$ = 190 K corresponds to the same $T$ at which the structural relaxation time $\tau_\alpha$ extrapolates to ∞ in the Vogel-Fulcher-Tammann (VFT) equation [19-21], where the VFT equation is likewise fit to relaxation or diffusion data for ordinary GF liquids above $T_g$, but below the crossover temperature $T_c$ estimated below.[31,32] The $u$ subscript serves a reminder that this characteristic temperature is estimated from $<u^2>$ and we continue to use this notational designation below for the other characteristic temperatures of glass-formation.

Following Dudowicz et al. [31], we may tentatively estimate the glass transition temperature $T_g$ by a Lindemann estimate appropriate for a *fragile* glass-former, $<u^2>^{½} / \sigma \approx 0.15$, which for the present system corresponds to the estimate, $T_{g,u}$ = 500 K. We deem this estimate of $T_g$ to be "reasonable" based on the observation made below that $T$ dependence of relaxation and diffusion in our metallic glass model returns to being nearly Arrhenius fashion in the 'glass' state below this characteristic temperature. Notably, quasi-thermodynamic measurements, such as specific heat measurements, show little or no evidence of any thermodynamic 'feature' near their glass transition temperature. [33] Correspondingly, we take this return to Arrhenius relaxation to be defining characteristic of glass-formation broadly, although is normally impossible to prove this is an equilibrium or non-equilibrium phenomenon. Recent simulations by Sciortino have shown that particles interacting with patchy interactions can exist in an equilibrium amorphous state having a finite and slowly varying configurational entropy, a lower free energy than the corresponding crystal, and relaxation and diffusion dynamics is nearly perfectly Arrhenius, i.e., an equilibrium glass. [34] Since many polymeric, and other fluids of complex-shaped molecules have conformational complexity and associated packing frustration, we believe that they also in some cases likewise form true equilibrium glass materials so we should not always think of glass-formation as a kinetic phenomenon. An explicit example of a polymeric fluid exhibiting this type of behavior has recently found [35] for a particular class of polymer liquids in the lattice cluster theory of the thermodynamics of polymer melts, in conjunction with the Adam-Gibbs model of glass-formation linking the thermodynamics to the fluid dynamics. [31] Of course, the glass state in most metallic GF materials is clearly just a metastable condition.

The generalized entropy theory of glass-formation [31,36] indicates that there two distinct regimes of glass-formation, a high temperature regime and low $T$ regime and the transition temperature separating these regimes is termed the 'crossover temperature', $T_c$. As in our previous work on the Cu-Zr metallic glasses of different composition, we identify this crossover temperature by the occurrence sharp deviation of $<u^2>$ from a linear variation, indicating the onset of strongly anharmonic interparticle interactions. In ordinary GF liquids, $\tau_\alpha$ scales as a power-law, $\tau_\alpha \sim [(T - T_c)/T_c]^{-\gamma}$, in a limited $T$ range above the 'crossover temperature' $T_c$, and we check below if this phenomenology applies to a material exhibiting FS glass-formation. The common observation of this type power-law scaling at intermediate in water would suggest that this scaling 'feature' is general in GF liquids [37,38], but as in the VFT equation the $T$ range where this scaling relation holds is limited. Nonetheless, the determination of this characteristic temperatures $T_o$ and $T_c$ provide important reference $T$, even if no actual divergence occurs at these $T$.

Finally, we consider the estimate the onset temperature of glass-formation $T_A$ at which non-Arrhenius dynamics and other deviations from simple fluid dynamics emerges in the liquid dynamics from $<u^2>^{½}$ data. In our previous work on Cu-Zr metallic glasses, we found that would could get a good rough estimate of $T_A$ is based on a complementary Lindemann type of criterion introduced by LaViolette and Stillinger [31,39] for the instability of the liquid state to ordering. In particular, we estimate $T_A$ from the condition that $<u^2>^{½} / \sigma$ is about 9 times its value at $T_g$.

Dudowicz et al. [31] found that this simple criterion was remarkably consistent with $T_g$ estimates calculated through direct computation in the entropy theory of glass-formation even in the case of polymer materials having different architectures. Based on this criterion, we estimate $T_{A,u}$. In the next section, we estimate $T_A$ and $T_c$ by the transitional method of finding the $T$ at which $\tau_\alpha$ departs from being Arrhenius and the power-law scaling near the crossover temperature $T_c$ noted above. Notably, the characteristic temperature $T_c$ is precisely defined as the inflection point temperature of $S_c T$ where $S_c$ is the configurational (i.e., non-vibrational) entropy of the fluid, [31] and thus can be calculated with precision from this model. We also consider this interpretation of $T_c$ below through the calculation of the excess entropy density [31] of our fluid relative to the glass state ($T < 450$ K).

**B. Precise Determination of the Onset Temperature $T_A$ of Non-Arrhenius Relaxation**

Following our previous work on the dynamics of Ni grain boundaries[29], the VFT temperature $T_0$ is estimated as the $T$ at which $<u^2>$ extrapolates to zero. The onset temperature $T_A$ for the onset of the glass transition regime was determined based on the entropy theory of glass-formation [31] that the (differential) activation energy $\Delta H$ in the high $T$ regime follows a universal quadratic $T$ dependence, $\Delta H(T) / \Delta H(T_A) = 1 + C [T / T_A - 1]^2$, where $\Delta H$ is specifically the differential activation energy defined by the local slope of the Arrhenius curve for $\tau_\alpha$ and the constant $C$ provides a measure of 'fragility' in the high $T$ range of glass-formation in which the VFT equation no longer applies. [31] In the present system we find $C$ to equal, $C = 23.4$. We mention that this $T$ dependence of the activation energy is essentially equivalent to that of the 'parabolic model' of Elmatad et al. [40,41] We emphasize that applicability of this relationship for the activation energy in the generalized entropy theory is limited to the high $T$ regime of glass-formation defined precisely by the $T$ range, $T_A > T > T_c$. [31] Based on the data summarized in Fig.

S5 we estimate $T_A$ to equal, $T_A = 927$ K, which is reasonably consistent with the rough estimate of $T_{A,u} = 906$ K, from $<u^2>$ in the previous section. Our previous study of Cu-Zr metallic GF liquids having a wide range of composition also exhibited a remarkably close correspondence between two rather different methods of estimating $T_A$. [3] Sastry and coworkers [42,43] have suggested an alternative method of estimating $T_A$ from the difference between the fluid entropy relative to the gas state and the pair correlation function estimate of this same quantity. $T_A$ is precisely prescribed in the generalized entropy theory [31] by a maximum in the entropy density (See Sect. E of SI). In Fig. 11 of the main text, we found that $T_A$ could also be estimated by the $T$ at which $<u^2>$ of the mobile particles becomes equal to $<u^2>$ of all the particles in the system.

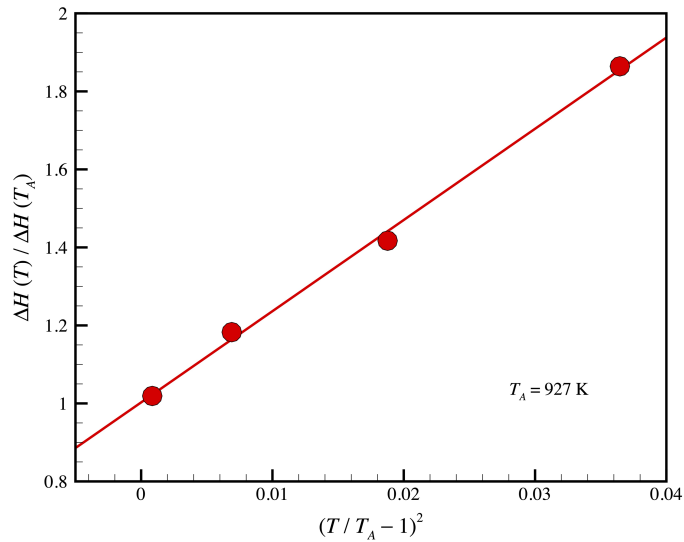

**Figure S5.** Activation energy as a function of temperature. Reduced apparent activation energy, $\Delta H(T) / \Delta H(T_A)$, for diffusion of $Al_{90} Sm_{10}$ metallic glass, as a function of reduced temperature, $\Delta H(T) / \Delta H(T_A) = C [(T - T_A)/ T_A]^2$, which allows us to estimate the onset temperature for our SM-Al metallic glass fluid $T_A = 927$ K, where the constant $C$ is a measure of fragility. [31] For the present system, $C = 23.4$. We previously applied this method of estimating $T_A$ associated with the grain boundaries[29] of Ni and the dynamics of polymeric GF liquids. [17,44]

## C. Determination of the Crossover Temperature $T_c$ Separating the High and Low Temperature Regimes of Glass-Formation

The crossover temperature $T_c$, defining the transition from a high and low $T$ regime of glass-formation [44] is estimated by fitting $\tau_\alpha$ from the intermediate scattering function to a power-law in the reduced temperature, $(T - T_c)/T_c$, resulting in an estimate of $T_c = 660$ K. This is to be compared to our rough estimate of $T_c$ from $<u^2>$ above, $T_{c,u.} \approx 650$ K. In our earlier work on Cu-Zr metallic glasses, [3] we found that $<u^2>$ consistently gave a reasonable rough estimate of this characteristic temperature.

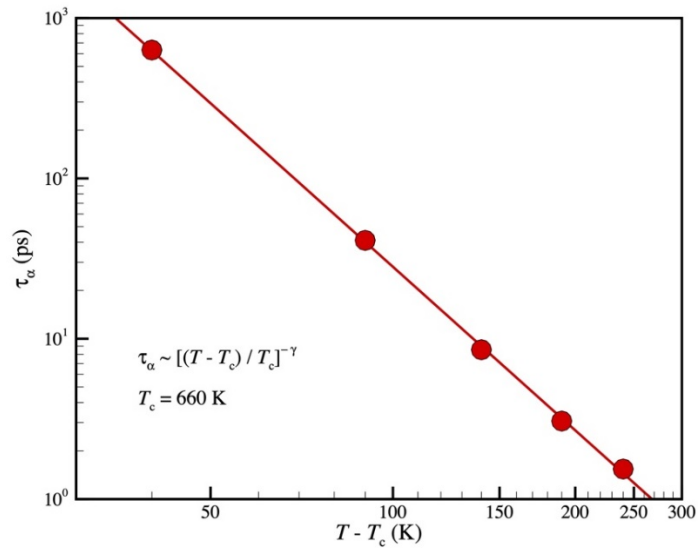

**Figure S6.** Estimation of the 'crossover tempearture', $T_c$. Near the crossover temperature $T_c$, sometimes termed the 'mode coupling temperature', the structural relaxation time $\tau_\alpha$ obtained from the intermediate scattering function normally exhibits a power-law scaling with $(T - T_c)/T_c$ to a good approximation. This type of power-law scaling is prevalently observed over a wide $T$ range in water, a prototypical FS glass-forming material.

**D. Attempt to Estimate $T_g$ and $T_o$ from VFT Equation**

The Vogel–Fulcher–Tammann (VFT) has often been used to define $T_g$ and $T_o$, but as noted above, the applicability of this expression is highly uncertain for fluids undergoing a fragile to strong transition. Strictly speaking, the VFT relation only holds as a good approximation for ordinary GF liquids for $T_c > T > T_g$. If we formally fit our relaxation time data for the Sm-Al metallic glass to the VFT equation[45], $\tau_\alpha / \tau_0 = \exp[D_f T_0 / (T - T_0)]$ we may be estimated the $T$ at which $\tau_\alpha$ formally diverges $T_o$ and $T_g$ may be precisely, if somewhat arbitrarily, be defined as the $T$ at which $\tau_\alpha$ extrapolates to 100 s. It is evident from Fig. S7 that the VFT equation provides a poor description of our relation data for $T$ greater than our estimates of $T_c$ above, and even if we attempt fitting our $\tau_\alpha$ data to the VFT in a higher $T$ regime, we estimate $T_g$ to equal $T_{g,VFT} = 603$ K and $T_{o,VFT} = 565$ K (See Fig. S7). Such characteristic temperature estimate is evidently questionable for systems undergoing a FS transition so that we may understand how spurious estimates of these $T$ may be obtained by blindingly applying the standard phenomenology of GF liquids to this rather distinct class of fluids.

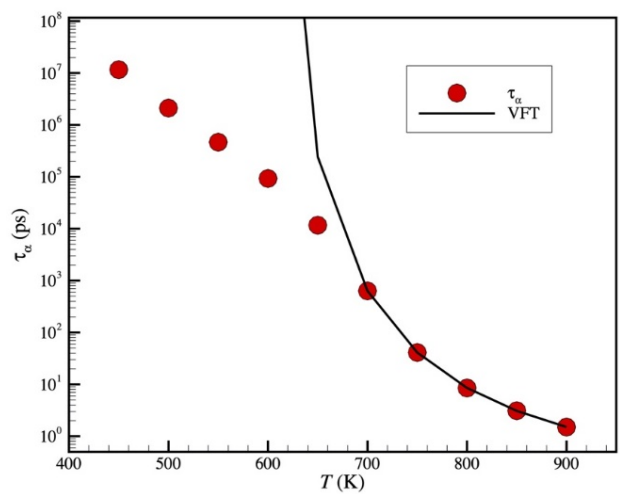

**Figure S7.** Attempt to estimate the glass transition temperature using the VFT relation. [19-21] This relation is clearly inapplicable beyond a rather limited $T$ range.

**E. Determination of the Lambda Transition Temperature, $T_\lambda$**

Glass-forming liquids exhibiting a FS transition characteristically exhibit true thermodynamic "anomalies" that are not observed in "ordinary" GF liquids. It is not clear whether this is an intrinsic difference between this class of GF liquids and ordinary GF liquids, or whether this difference relates the practical observability of these thermodynamic features due to variations in the magnitude of the effects observed or the accessibility of the $T$ range over which these thermodynamic features arise. One of the most conspicuous features observed in materials exhibiting a FS transition is conspicuous peak in the specific heat $C_p$ of a genuine thermodynamic origin rather than just a signature of going out of equilibrium, as found in measurements in ordinary GF liquids and a feature essentially absent for GF liquids undergoing a FS transition. [46] In the well-studied case of water, a number of thermodynamic and transport properties are observed [47-49] to vary strongly with $T$ upon approaching a 'singular temperature' $T_s$ from above, and this $T$ was designated the 'lambda temperature' $T_\lambda$ by Angell and coworkers [33] because of its characteristic singular shape, reminiscent of other materials undergoing second order or near second order phase transitions. Angell and coworkers particularly emphasize the analogy with the polymerization of sulfur, which may be described as equilibrium polymerization transition or mathematically as an Ising model with an applied field., the field known to provide a source of rounding in this type of second order transition. This rounded thermodynamic transition point of view of glass-formation has been amplified upon by Douglas et al. [50] and provides a precise mathematical framework for modeling the dynamic heterogeneity observed in GF liquids. [51,52]

Simulations of water have also indicated that the specific heat $C_p$ and isothermal compressibility, [53,54] exhibit a maximum at a common characteristic temperature, which can be taken as a definition of $T_\lambda$ and Fig. S8 shows that this same pattern of behavior arises in our Sm-Al metallic glass, as expected, where we see that $C_p$ and the thermal expansion coefficient both exhibit an extremum near $T = 750$ K. We also examined the 4-point density correlation function $\chi_4$ as a function and the noise exponent governing potential energy fluctuations, which likewise have a peak near this characteristic temperature, as seen before in simulations of water [53,54] Our simulation estimates of these 'linear response' properties exhibit a remarkably similar trend to water, except from the important matter that the position of the peak in $C_p$ arises near $T_c$ in water, while $T_\lambda$ occurs well above $T_c$ in our Sm-Al metallic GF material; this situation greatly helps in the investigation of these thermodynamic 'anomalies' in the metallic glass thermodynamic properties. There are other materials for which $T_\lambda$ lies below $T_c$ and other fluids in which it is well above $T_c$. For example, we note that Monte Carlo simulations of the binary Lennard-Jones model indicate a sharp peak in the equilibrium specific heat at a $T$ well below $T_c$, [55] making this thermodynamic feature inaccessible by equilibrium molecular dynamics simulation because of the low $T$ at which this feature arises and the corresponding long relaxation times at such temperatures. The case of $BeF_2$, on the other hand is a fluid in which this peaking in the specific heat, compressibility and other properties occurs at very high $T$ so that this transition is practically inaccessible because of the high T range involved. [46,56-58] In the case of water, the $T$ at which the $C_p$ peaks is apparently rather close to $T_c$.[29, 30] Evidently, these extrema in thermodynamic properties occur with some generality in GF liquids, but the location of these thermodynamic 'anomalies', relative to the characteristic temperatures of glass-formation, shows significant variability. It is noted that specific heat anomalies having the superficial appearance of thermodynamic transitions have been observed in metallic glass materials [59] where the resemblance to the $C_p$ 'anomaly' in water has been noted [33,48,59] and striking lambda transitions resembling the lambda transition in He[4] near its superfluid transition has been observed in other GF materials. [60,61] This is clearly some kind of thermodynamic transition phenomena, even if no true phase transition is involved.

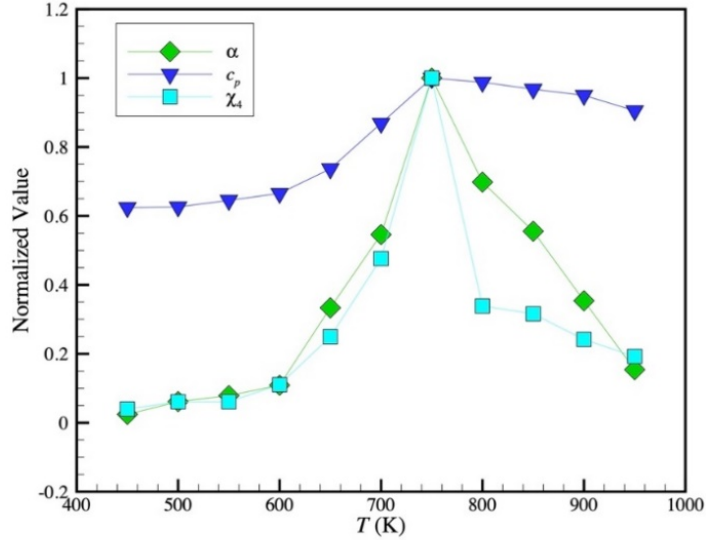

**Figure S8.** Comparison of basic thermodynamic response functions and colored noise exponent of potential energy fluctuations. Specific heat, peak value of 4-point density function $\chi_4$ and potential energy noise exponent (see our companion paper focussing on the fast dynamics of Sm-Al [62] for explanation of this data) as function of $T$ in a simulated Sm-Al metallic glass. We see from Fig. 21 b of the main text that the isothermal compressibility exhibits a small maximum at the same $T$ as $C_p$ (see also Fig. 19 of the main text where $C_p$ is plotted in a way that its peak is more prominent. A peak in the intensity of noise associated with mean square coordination fluctuations, which should be reflected in potential energy fluctuations, has been observed in previous simulations of water. [53] We see that all these properties exhibit a peak near a common temperature near 750 K. We define the temperature at which this specific maximum occurs to be the 'lambda temperature', $T_\lambda$. We note that such a T also arises in superionic crystalline materials and our notation for this transition follows that commonly used in this class of materials. [63]

We may gain further insight into $T_\lambda$ by considering the $T$ dependence of the excess entropy $S_{exc}$ of our GF liquid relative to the entropy of the fluid in its glass state, a quantity that has historically been considered as an estimate of the configurational entropy of a fluid, $S_c$. This is a central thermodynamic property in relation to the dynamics of GF liquids. Metallic glass materials

19are apparently particularly favorable systems for such an estimation of $S_c$ since the vibrational entropy has been found to relatively small in such systems.[64] The generalized entropy theory [31] prescribes that when considering physically realistic compressible materials the quantity that should be considered in relation to dynamics is the $S_c$ divided by the material volume, the configurational entropy density (Apart from the theoretical rationale, the activation energy does not recover Arrhenius relaxation at elevated $T$ if entropy per unit mass is considered.). Our calculation of the excess entropy density closely follows our calculation of this quantity in the case of superionic $UO_2$ except we utilize the entropy of our metallic GF material in its glass state rather than the entropy of the crystal in its low $T$ harmonic crystal state below an onset temperature $T_\alpha$, somewhat analogous to $T_o$ for our GF liquid where complex dynamics first emerges in the crystalline material. [63] We crudely estimate $S_c$ by the excess entropy $S_{exc}$, the difference between the total material entropy and its entropy in the glass state ($T < 450$ K), where the total entropy $S$ is estimated based on a determination of the specific heat and the $S_g$ in glass state is estimated below the onset temperature, $T_\alpha$. The difference of $S$ minus its value below $T_\alpha$ defines an "excess entropy" $S_{exc}$, and we further normalize $S_{exc}$ by the material volume $V$ so we obtain the excess entropy density.

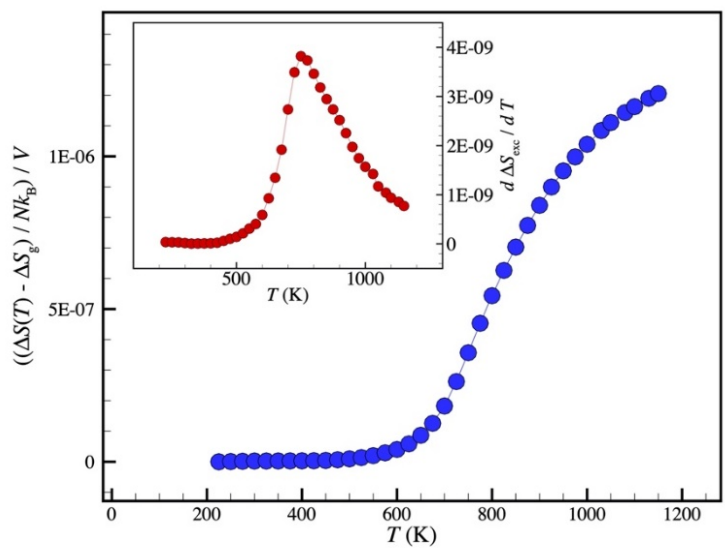

**Figure S9.** Estimation of excess entropy density as a function of temperature of a Sm-Al metallic GF material undergoing a FS transition. Inset shows the derivative of the excess entropy density. Note that the excess entropy density is defined relative to its value estimated over a range of $T$ in the glass state. The derivative of the excess energy density with respect to $T$ shows a maximum at $T = 750$ K, which coincides with $T_\lambda$ determined by other measures to within numerical uncertainty determined by other measures.

We note that several studies [65-67] of metallic GF liquids have successively applied the Adam-Gibbs model to describe diffusion coefficient data in a semi-quantitative approximation [68], but we avoid this procedure until we have a better method of estimating $S_c$ from our metallic glass data. The common procedure of subtracting the crystal entropy $S_{xstl}$ from $S$ of the GF liquid to estimate $S_c$ cannot be applied to our metallic GF material because the crystalline form of this material is not available. In previous simulations of superionic $UO_2$, we found that the AG model with Sc estimated from the excess entropy in comparison to the crystal provided a rather good estimate of the $T$ dependence of the activation energy for O diffusion in this material[63], in accord with previous experimental studies successfully applying the AG model to this broad class of anharmonic *crystalline* materials (See Zhang et al. [63] for a discussion).

**F. Fragile-Strong Transition Temperature, $T_{FS}$**

The FS transition in GF liquids is signaled by change in the differential activation energy with $T$. In particular, the sharp increase in this apparent activation energy, i.e., the magnitude of the local slope of the Arrhenius curve for diffusion or relaxation, first increases upon lowering $T$ below $T_A$, but this slope peaks, and then falls as the system is cooled further towards the glass state where Arrhenius relaxation and diffusion re-emerge. The FS transition is thus a transition between a high and low $T$ Arrhenius regimes, each having their own distinct activation energies and it is natural to define the intermediate T at which the differential activation energy peaks, the FS transition temperature, $T_{FS}$. Notably, a peak in the differential activation energy is observed rather generally in superionic crystalline materials (e.g., $SrCl_2$, $CFl_2$, $SrF_2$, $BaF_2$, $UO_2$) [59,63] at a $T$ somewhat lower than $T_\lambda$, observations which are striking similar to our Sm-Al metallic glass.

We illustrate the $T$ dependence of the differential activation energy $E_{\text{diff}}(T)$ in Fig. S10, where we identify the maximum in this curve with $T_{FS}$, yielding $T_{FS} = 700$ K. This $T$ is evidently intermediate between $T_c$ and $T_\lambda$ in our Sm-Al metallic GF material. We also clearly observe the initial increase of $E_{\text{diff}}(T)$ upon cooling nearly coincides with $T_A$ and that $E_{\text{diff}}(T)$ saturates to a constant value near 500 K, a $T$ close to our estimates of $T_g$ above. Thus, $T_A$ and $T_\lambda$ demark the beginning, middle of the glass-transition as function of $T$ and the $T$ at which $E_{\text{diff}}(T)$ saturates to a constant at low $T$ nearly coincides with our estimate of $T_g$ from $<u^2>$, based on the application of a Lindemann criterion, $T_{g,u} = 500$ K. An empirical correlation [69] for numerous systems exhibiting FS glass-formation has indicated the correlation $T_{FS} \approx 1.36\ T_g$ or about 680 K, in good qualitative consistency with our estimate above, $T_{FS} \approx 700$ K. Of course, non-equilibrium effects probably contribute to this slowing down of the $T$ dependence of $E_{\text{diff}}(T)$ at low $T$, as in ordinary

GF liquids. Nonetheless, we think the estimation of $T_g$ from *either* $E_{diff}(T)$ or $<u^2>$ seem to offer a practical and remarkably effective means of estimating a physically meaningful $T_g$ and other characteristic temperatures for materials undergoing a FS type of glass-formation.

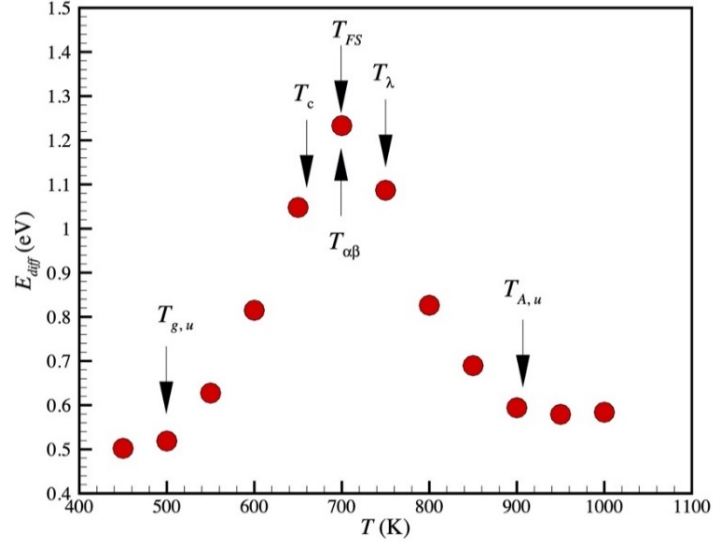

**Figure S10.** Differential activation energy $E_{diff}$ obtained from the diffusion coefficient of an $Al_{90}Sm_{10}$ metallic glass-forming material.

### G. Estimation of the Bifurcation Temperature, $T_{\alpha\beta}$

It is commonly observed that an extrapolation of the Arrhenius curve for the Johari-Goldstein relaxation time $\tau_{JG}$ intersects the curve describing the α-relaxation time $\tau_\alpha$ at a temperature that is near the crossover temperature $T_c$, [70,71] but there are exceptions to this empirical rule in some GF liquids. [72] We may estimate this αβ 'bifurcation temperature' $T_{\alpha\beta}$ from our estimates of $\tau_{JG}$ and $\tau_\alpha$ to determine whether this common phenomenology applies to our metallic glass system exhibiting FS glass-formation. In Figure S11, we see that the fitted curves for these relaxation times intersect at 700 K, which is reasonably close to our estimate of $T_c$ above for this material, $T_c = 660$ K. Note that it is possible that $\tau_{JG}$ deviates from an Arrhenius variation at higher $T$ than the data indicated so this intersection point relies on an extrapolation.

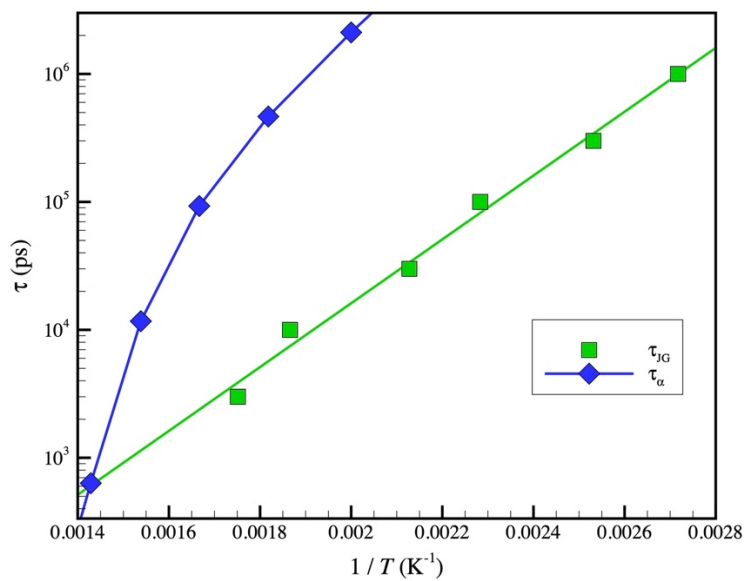

**Figure S11.** Estimation of the extrapolate α-β 'bifurcation temperature', $T_{\alpha\beta}$. The extrapolation of $\tau_{JG}$ from low $T$ intersects with $\tau_\alpha$ at $T = 700$ K, which defines the 'bifurcation temperature', $T_{\alpha\beta}$. Apparently, this characteristic temperature is close to $T_c = 660$ K of the $Al_{90}Sm_{10}$ metallic GF liquid. This plot is also shown in the main text, but we reproduce this plot here for completeness of the characteristic temperature characterization.

## 4. Quantifying Immobile Particle Cluster Size

Previous work has shown that the lifetime of the immobile particle clusters scales consistently with the average size (number of particles) in the immobile particle clusters for both polymeric and metallic GF liquids,[73,74] and we find this relation continues to hold for our metallic GF liquid for the $T$ range in Figure S12, i.e., $T \geq 700$ K. See Starr et al. [73] and Wang et al. [74] for an extended discussion of different types of dynamic heterogeneity in polymeric and metallic GF liquids where it is found that the structural characteristics of these heterogeneities are remarkably similar in these chemically rather distinct materials.

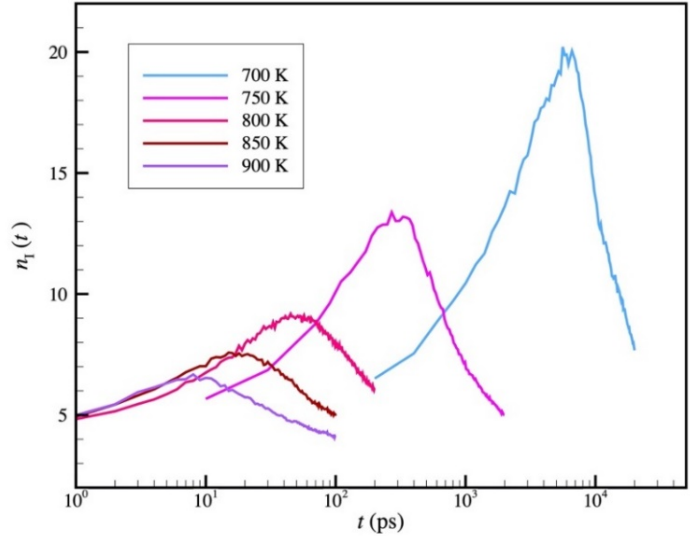

**Figure S12.** Time evolution of the Immobile particle cluster size as a function of time and temperature. Immobile cluster size in an Al$_{90}$ Sm$_{10}$ metallic glass as a function of time for $T \geq 700$ K. The immobile particles are defined as the 5 % least mobile particles in the system at certain time window.

**Table S1. Summary of characteristic temperature estimates by different methods**

| $T_{0, u}$ | $T_{0,VFT}$ | $T_{g, u}$ | $T_{g,VFT}$ | $T_{c,u}$ | $T_c$ | $T_{\alpha\beta}$ | $T_{FS}$ | $T_\lambda$ | $T_{A, u}$ | $T_A$ |
|---|---|---|---|---|---|---|---|---|---|---|
| 190 | 565 | 500 | 603 | 650 | 660 | 700 | 700 | 750 | 906 | 927 |

## 5. Interpretation of $\tau_\alpha$ as a 'Structural Relaxation Time'

The relaxation time $\tau_\alpha$ of the intermediate scattering function corresponding to a scale on the order of the interatomic distance is often designated as being a structural relaxation time with the implicit understanding that this quantity bears an approximate relationship with the shear stress relaxation time, the measure of the structural relaxation most favored by experimentalists. It is generally appreciated that this correspondence is not an exact relationship, but it is nonetheless generally assumed that such a relation holds qualitatively so the designation of $\tau_\alpha$ as a 'structural relaxation time' is still appropriate. In this section, we show evidence for this interrelation in the case of a Cu-Zr that we have recently studied and which we think is sufficient to establish our qualitative point that these relaxation times indeed 'track' each other (see Fig. S12). On the other hand, these calculations also show that these quantities cannot be exactly identified. A recent study by Jaiswal et al. [75] also show that the structural relaxation time $\tau_\alpha$ exhibits a strikingly linear relationship with shear stress relaxation time in a ternary $Cu_{40}Zr_{51}Al_9$ metallic liquid system [See Fig. 9(c) of Ref. [75]]. This scaling relation between the shear stress relation time and $\tau_\alpha$ should be checked in other GF liquids.

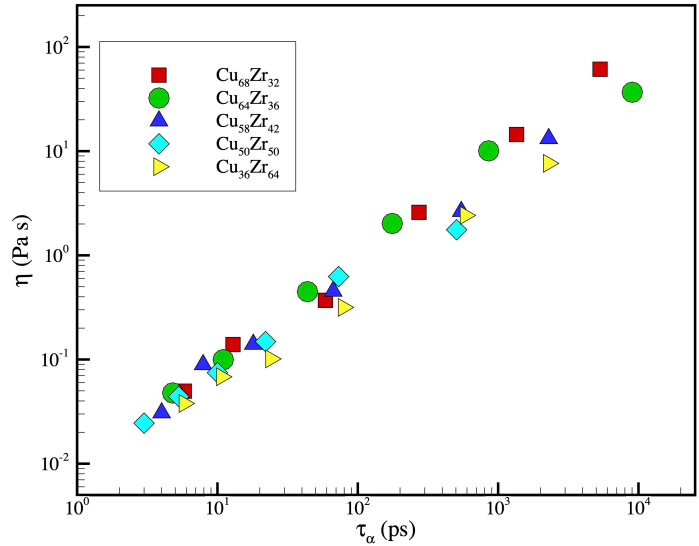

**Figure S13.** Comparison between viscosity and structural relaxation times of Cu-Zr alloys with five different compositions. [30]

## 6. Apparent Exponent in Decoupling Relations

**Table S2. Summary of exponent between different characteristic times and diffusivity**

| Relation | $t^* \sim \tau_\alpha^{1-\zeta^*}$ | $\tau_M \sim \tau_\alpha^{1-\zeta M}$ | $D/T \sim (1/\tau_\alpha)^{1-\zeta\alpha}$ | $D/T \sim (1/\tau_{JG})^{1-\zeta JG}$ | $D/T \sim (1/t^*)^{1-\zeta^*}$ |
|---|---|---|---|---|---|
| Exponent | 0.54 | 0.36 | 0.43 | -0.06 | -0.23 |